\documentclass[12pt,a4paper]{article}
\usepackage{jheppub}
\usepackage{multirow}

\usepackage{bbold}
\usepackage{amsmath,amssymb,amsfonts}
\usepackage[footnotesize]{caption}
\usepackage{braket}
\usepackage{mathtools}
\usepackage{url}
\usepackage[linkcolor=blue,citecolor=blue,urlcolor=blue]{hyperref} 

\usepackage{cleveref}
\crefname{section}{Sec.}{Secs.}
\Crefname{section}{Section}{Sections}
\crefname{table}{Tab.}{Tabs.}
\newcommand{\Figref}[1]{Fig.~\ref{#1}}
\newcommand{\Tabref}[1]{Tab.~\ref{#1}}
\newcommand{\eq}[1]{Eq.~\eqref{#1}}

\allowdisplaybreaks

\DeclareMathOperator{\im}{Im}

\newcommand{\MeV}{\:\text{MeV}}
\newcommand{\GeV}{\:\text{GeV}}

\newcommand{\Fmc}{\mathit{c}} 
\newcommand{\Y}{\mathcal{Y}}
\newcommand{\QL}{Q_L}
\newcommand{\LL}{L_L}

\newcommand{\ckm}{\textnormal{CKM}}
\newcommand{\muEW}{{\mu_\text{ew}}}

\def\beq{\begin{equation}}
\def\eeq{\end{equation}}
\def\bea{\begin{eqnarray} }
\def\eea{ \end{eqnarray} } 
\newcommand{\nn}{\nonumber}

\makeatletter
\def\bstctlcite{\@ifnextchar[{\@bstctlcite}{\@bstctlcite[@auxout]}}
\def\@bstctlcite[#1]#2{\@bsphack
 \@for\@citeb:=#2\do{%
   \edef\@citeb{\expandafter\@firstofone\@citeb}%
   \if@filesw\immediate\write\csname #1\endcsname{\string\citation{\@citeb}}\fi}%
 \@esphack}
\makeatother

\title{An Unfamiliar Way to Generate the Hierarchy of Standard Model Fermion Masses}
\author[a,b]{S.~Baek}
\author[b,c,d]{J.~Kersten}
\author[b,e]{P.~Ko}
\author[f,g]{L.~Velasco-Sevilla}

\affiliation[a]{
Department of Physics, Korea University, Seoul 01841, South Korea}
\affiliation[b]{
School of Physics, Korea Institute for Advanced Study, 
Seoul 02455, South Korea}
\affiliation[c]{
Department of Physics and Technology, University of Bergen,
5020 Bergen, Norway}
\affiliation[d]{Department of Physics, Yonsei University, Seoul 03722,
South Korea}
\affiliation[e]{Quantum Universe Center, Korea Institute for Advanced Study, 
Seoul 02455, South Korea}
\affiliation[f]{
Center for Quantum Spacetime, Sogang University, Seoul 121-742, South Korea}
\affiliation[g]{
Department of Physics, Sogang University, Seoul 121-742, South Korea}

\emailAdd{sbaek1560@gmail.com}
\emailAdd{joern.kersten@uib.no}
\emailAdd{pko@kias.re.kr}
\emailAdd{lilianak@sogang.ac.kr}

\abstract{While the properties of the observed Higgs boson
agree with the Standard Model predictions, the hierarchy of
fermion masses lacks 
an explanation within the model.
In this work, we consider a fresh approach to this problem, involving a different Higgs doublet responsible for each quark mass.
We construct a model with a gauged, non-anomalous $U(1)$ family symmetry
that fixes which fermion couples to which doublet with an $\mathcal{O}(1)$ Yukawa coupling.
The hierarchy of masses is 
generated by the hierarchy of vacuum expectation values of the Higgs fields.
The model generically predicts a light, weakly coupled pseudoscalar.
We verify that the model satisfies constraints from flavour changing
neutral currents, Higgs phenomenology and electroweak precision tests.
}

\arxivnumber{2309.07788}

\begin{document}
\begin{flushright}
    {CQUeST-2023-0724 \\ KIAS-P23038}
\end{flushright}

\maketitle

\section{Introduction}

In the Standard Model (SM) the electroweak (EW) gauge symmetry is broken
by a single Higgs doublet of scalar fields.  The vacuum expectation value
(vev), $v/\sqrt{2}$, of the neutral component can explain the appearance of
fermion and gauge boson masses.
The properties of the observed Higgs boson with a mass of $125\GeV$
agree with the SM predictions.  However, it would be too early to exclude additional Higgs bosons beyond the SM\@. 
To motivate the essence of this work,
consider the top Yukawa coupling $y_t \, h \, \overline{t_R} \, t_L$.
Once the Higgs field acquires a vev, a top mass
$m_t = y_t\,v/\sqrt{2}\simeq y_t \cdot 174\GeV$ arises.  Experimentally,
$m_t \simeq 173\GeV$, and thus the Yukawa coupling $y_t$ is an $O(1)$
number.  In contrast, the Yukawa coupling of the up quark is of order
$10^{-5}$ in the SM.

The origin of this hierarchy has been tackled in many ways,
for example, by accidental flavor models \cite{Ferretti:2006df}, radiative mass models \cite{Babu:1989fg}, warped compactification \cite{Carena:2004zn,Archer:2012qa}, partial compositeness \cite{Goertz:2023nii}, clockwork mechanisms \cite{Patel:2017pct}, and modular symmetries \cite{King:2020qaj,Ding:2022bzs,Ding:2022aoe}.  One
widely-used approach employs family symmetries together with the
Froggatt-Nielsen mechanism \cite{Froggatt:1978nt}.  This mechanism uses
additional scalars (flavons), $\theta_i$, which do not contribute to EW
symmetry breaking but acquire vevs at some high scale, $M$, such that
the Yukawa couplings of the SM fermions are given by
$(\braket{\theta_i}\!/M)^p$, with $p$ determined by the charges of a
gauged family symmetry.
For a sufficiently small ratio $\braket{\theta_i}\!/M$
(or large enough $p$), this results in the observed hierarchy of Yukawa
couplings.  However, only in a few studies, e.g., \cite{Tsumura:2009yf,Berger:2014gga,Bauer:2016rxs,Khoze:2017tjt,Li:2019bcr,King:2020mau,Koivunen:2023led}, the production of flavons in high-energy
experiments has been discussed.  The vast majority of models aim for
predictions of hierarchies and mixings, but not for the prospects of
detecting flavons since the breaking scales of the symmetries involved
are very high. We therefore have no direct experimental test of these
models.
As an alternative, less-explored route and in view of the basic observation
that there are three families of fermions, we pursue the idea that
there are also three families of Higgs fields, each containing one
doublet with hypercharge $\mathcal{Y}=\frac{1}{2}$ and one doublet with
$\mathcal{Y}=-\frac{1}{2}$.  The doublets couple to the SM fermions such that the
hierarchy of their vevs controls the hierarchy of fermion masses with
all Yukawa couplings of $\mathcal{O}(1)$.  Thus, we postulate
\begin{equation}
	m_d \sim v_1 \quad,\quad
	m_u \sim v_2 \quad,\quad
	m_s \sim v_3 \quad,\quad
	m_c \sim v_4 \quad,\quad
	m_b \sim v_5 \quad,\quad
	m_t \sim v_6
\end{equation}
and
\begin{equation} \label{eq:vevhierarchy}
	v_1 \sim v_2 \ll v_3 \ll v_4 \ll v_5 \ll v_6 \,.
\end{equation}
Multi-Higgs models with this motivation 
\cite{Escudero:2005hk,Escudero:2005ku,Hill:2019cce, Hill:2019ldq,Altmannshofer:2021hfu}
or with the idea that each fermion has its corresponding Higgs
\cite{Porto:2007ed,Porto:2008hb,BenTov:2012xp,BenTov:2012cx}
have been considered in the past.
In particular, ref.~\cite{Escudero:2005hk} presented a supersymmetric model
with six doublets and analyzed the EW breaking as well as the most
important constraints from flavour-changing neutral currents.
The structure of the Yukawa couplings relied on a Fritzsch texture
\cite{Fritzsch:1977vd}.  We take the work of ref.~\cite{Escudero:2005hk} 
as a basis but consider a non-supersymmetric model with a gauged family
symmetry that yields the required form of the fermion mass matrices. Specifically, we employ a $U(1)_F$ extension of the SM gauge group that
is family-dependent.  Its role is not to fix the Yukawa coupling
hierarchies but rather to determine the textures of the Yukawa
matrices.  Having a gauged family symmetry forces us to ensure the
cancellation of all gauge anomalies, which restricts the assignment of
$U(1)_F$ charges and thus reduces the number of free parameters.

The structure of the work is as follows.  After presenting the $U(1)_F$
model, we discuss the cancellation of anomalies in \cref{sec:details}.
In \cref{sec:Yukawas}, we investigate how to obtain Yukawa matrices that
are consistent with the observed quark masses and mixings, arriving at a
concrete realization of the model that is analyzed in detail in the
following sections.  First, we discuss the scalar potential in
\cref{sec:scalarpotential} and show that it is possible to find
parameters leading to successful EW symmetry breaking and a realistic
scalar mass spectrum.  Second, we consider the constraints from
flavour-changing neutral currents and CP violation in \cref{sec:fcnc}.
Third, we determine the changes of EW precision observables and discuss
whether the anomalous value of the $W$ boson mass reported by the CDF
experiment \cite{CDF:2022hxs} can be accommodated in the model.
Finally, we conclude in \cref{sec:conclusions}.

\section{Anomaly Cancellation and Matter Content \label{sec:details}}

\subsection{General Model}

We aim for building a model for which the necessity of any particular fermion comes with a reason. In this respect, we first aim at finding solutions where only the SM fermions are present. We will then talk about how other exotic fermions can enter the picture and the purpose they can serve. 
Our model is like the SM plus the additional Higgs doublets, singlet scalars and the additional Abelian gauge boson corresponding to $U(1)_F$. For clarity of the presentation we write the terms in the Lagrangian that are modified with respect to the SM Lagrangian
\begin{equation}
\mathcal{L}=\sum_{f} \, i\bar\psi_f \gamma^\mu D_\mu \psi_f + \sum_{m=1}^{6} \left(D^\mu H_m\right)^\dagger D_\mu H_m +
\sum_{n=1}^3 \left(D^\mu \phi_n\right)^\dagger D_\mu \phi_n -
V(H_m,\phi_n) + \mathcal{L}_\text{Yuk} \,,
\end{equation}
where $V(H_m,\phi_n)$ is the Higgs potential involving Higgs doublets $H_m$
and singlets $\phi_n$, which we will specify in \cref{sec:scalarpotential}.
The mass terms of the SM fermions come from $\mathcal{L}_\text{Yuk}$
given below in \eq{eq:MatterLagrangian}.
The action of the covariant derivative $D_\mu$ on the Higgs doublets $H_m$,
the singlets $\phi_n$, and the SM fermions $\psi_f$ is respectively given by
\begin{align}
D_\mu H_m &= \left(\partial_\mu -i g_2  W_\mu(x)\cdot \tau -i g_1 \mathcal{Y}_{H_m} \, B_\mu(x)  -ig_F \, c_{H_m} Z'_{\mu}(x) \right)H_m \,,
\nonumber\\
D_\mu \phi_n &= \left(\partial_\mu- ig_1 \, \mathcal{Y}_{\phi_n} B_\mu(x) -ig_F \, c_{\phi_n} Z'_{\mu}(x) \right) \phi_n \,,
\nonumber\\
D_\mu \psi_{fL} &= \left(\partial_\mu - ig_3 k\, G_{\mu a} \lambda_a -ig_2 W_\mu(x)\cdot\tau - ig_1 \, \mathcal{Y}_{\psi_{fL}} B_\mu(x) -ig_F \, c_{\psi_{fL}} Z'_{\mu}(x) \right) \psi_{fL} \,,
\nonumber\\
D_\mu \psi_{fR} &= \left(\partial_\mu -i g_3 k\, G_{\mu a} \lambda_a - ig_1  \mathcal{Y}_{\psi_{fR}}\, B_\mu(x) -ig_F \, c_{\psi_{fR}} Z'_{\mu}(x) \right) \psi_{fR} \,,
\end{align}
where $\tau^i=(1/2)\, \sigma^i$ are the $SU(2)_L$ generators.
Besides, $\mathcal{Y}_{H_m}$, $\mathcal{Y}_{\phi_n}$,
$\mathcal{Y}_{\psi_{fL}}$, and $\mathcal{Y}_{\psi_{fR}}$ are the
hypercharges of the scalar doublets and singlets
as well as the left- and handed-fermions, see \Tabref{tab:SMQN}.
Of course gluons, $G_{\mu a}$, only couple to quarks, so $k=0$ for
leptons and $k=1/2$ for quarks. The new gauge interaction, with boson
$Z'_\mu(x)$ and coupling $g_F$, fixes the charges $c_{\psi_{fL,R}}$, see
also \Tabref{tab:SMQN}, of the fermions of the SM according to anomaly cancellation conditions.

\begin{table}
\centering
\renewcommand{\arraystretch}{1.2}
\begin{tabular}{|c|ccccccc|}
\hline
 & $u_{L\, i}$ & $u_{R\,i}$ & $d_{L\, i}$ & $d_{R\, i}$ & $\nu_{L\, i}$ & $e_{L\, i}$ & $e_{R\, i}$ \\
\hline
 $Q_f$ & $\frac{2}{3}$ & $\frac{2}{3}$ & $-\frac{1}{3}$ & $-\frac{1}{3}$ &
 $0$ & $-1$ & $-1$  \\
 $\mathcal{Y}_f$ & $\frac{1}{6}$ & $\frac{2}{3}$ & $\frac{1}{6}$ &
 $-\frac{1}{3}$ & $-\frac{1}{2}$ & $ -\frac{1}{2}$ & $-1$  \\
 $I_3^f$ & $\frac{1}{2}$ & $0$ & $-\frac{1}{2}$ & $0$ & $\frac{1}{2}$ &
 $-\frac{1}{2}$ & $0$ \\
\hline
$U(1)_F$ & $\Fmc_{Q_{L\, i}}$ & $\Fmc_{u_{R\, i}}$ & $\Fmc_{Q_{L\, i}}$ & $\Fmc_{d_{R\, i}}$ & $\Fmc_{L_{L\, i}}$ & $\Fmc_{L_{L\, i}}$ & $\Fmc_{e_{R\, i}}$ \\
\hline
\end{tabular}
\caption{Quantum numbers of the model.\label{tab:SMQN}}
\end{table}

Anomaly cancellation of course does not restrict the scalar sector of the theory, but just as it happens with the families of the SM, we consider that the scalar sector contains three Higgs doublets with hypercharge $+1/2$ and three Higgs doublets with hypercharge $-1/2$, coupling to down quarks and up quarks, respectively
\begin{equation} \label{eq:HiggsDoublets}
H_{1(3,5)} = \begin{pmatrix}
H^+_{1 (3,5)}\\
H^0_{1 (3,5)}
\end{pmatrix}
\quad,\quad
H_{2(4,6)}= \begin{pmatrix}
H^0_{2 (4,6)}\\
H^-_{2 (4,6)}
\end{pmatrix} .
\end{equation}
As we will see later on in \cref{sec:scalarpotential}, in order to achieve a realistic scalar
mass spectrum, we need also scalar singlets. For the
particular model of \cref{sec:specificexample} we need three
singlets. The charges of this example are given in
\Tabref{tbl:ModA}. We decompose the neutral parts of the doublets in
\eq{eq:HiggsDoublets} in terms of component fields as
\begin{equation} \label{eq:H0Decomposition} 
H^0_l =
\frac{1}{\sqrt{2}} \, (v_l + \sigma_l + i \varphi_l) 
\quad,\quad
H^0_k =
\frac{1}{\sqrt{2}} \, (-v_k - \sigma_k + i \varphi_k) \,,
\end{equation}
where $l=1,3,5$ and $k=2,4,6$.
The minus signs in the decomposition of the $\mathcal{Y}=-1/2$ Higgs fields $H_k$
$(k=2,4,6)$ are introduced in such a way that the $\mathcal{Y}=1/2$ fields
\begin{equation}
\widetilde{H}_k = \epsilon H_k^* =
\begin{pmatrix}
H_k^+ \\
\frac{1}{\sqrt{2}} \, (v_k + \sigma_k + i \varphi_k)
\end{pmatrix}
\end{equation}
have the same sign convention as $H_l$ ($l=1,3,5$) 
(here $\epsilon$ is the anti-symmetric tensor in two dimensions, $\epsilon_{12}=-\epsilon_{21}=1$, $\epsilon_{ii}=0$).
In this way, we follow the sign convention in \cite{Grimus:2007if},
which enables us to use the results of that work to determine the EW
precision parameters in \cref{sec:T}.

The matter Lagrangian involving the fields
$Q_L$, $L_{L}$ (quark and lepton $SU(2)_L$ doublets, respectively),
$u_R$, $d_R$ (quark singlets) and $e_R$ (lepton singlets) is given by  
\begin{align}
- \mathcal{L}_\text{Yuk} &=
\overline{Q}_{Li}
\left[ (Y_1^d)_{ij}\, H_1+ (Y_3^d)_{ij}\, H_3 + (Y_5^d)_{ij}\, H_5 \right] d_{Rj}
\nonumber\\
&+
\overline{L}_{Li}
\left[ (Y_1^e)_{ij}\, H_1 + (Y_3^e)_{ij}\, H_3 + (Y_5^e)_{ij}\, H_5 \right] e_{Rj}
\nonumber\\
&+ \overline{Q}_{Li}
\left[ (Y_2^u)_{ij}\, H_2 + (Y_4^u)_{ij}\, H_4 + (Y_6^u)_{ij}\, H_6\right] u_{Rj}
 + \text{h.c.} \,,
\label{eq:MatterLagrangian}
\end{align}
where $Y_n^{q(e)}$ are the Yukawa matrices associated with each
Higgs field $H_n$.
We choose $U(1)_F$ charges such that couplings of the type
\begin{equation}
\label{eq:unwQcoups}
\overline{Q}_{L\, i} \,\epsilon H^*_k d_{R\, j} \quad,\quad 
\overline{Q}_{L\, i} \,\epsilon H^*_l u_{R\, j} \quad
{}_{l \, \in \, \{1,3,5\} \;,\; k\, \in\, \{2,4,6\}}
\end{equation}
are forbidden by the gauged family symmetry.%
\footnote{Distinguishing two Higgs doublets by additional
Higgs gauge symmetries was proposed as an alternative solution to the 
Higgs-mediated FCNC problems in 2HDM, generalizing the usual 
softly broken $Z_2$ symmetry \cite{Ko:2012hd}.}
Otherwise, $\braket{H_6}$ would contribute
equally to the bottom and top quark masses, which does not comply with  
our idea on the fermion mass hierarchies.

The electric charges are related to weak isospin and hypercharge by
\begin{equation}
Q_f = I_3^f + \mathcal{Y}_f \,.
\end{equation}

The cancellation conditions of triangle mixed anomalies, with external gauge boson lines and internal lines of a SM fermion, of the type $U(1)_{\rm{F}}-G^i_{\rm{SM}}-G^i_{\rm{SM}}$, where $G^i_{\rm{SM}}=U(1)_{\rm{Y}}$, $SU(2)_{\rm{L}}$, $SU(3)_{\rm{C}}$, are given by $A^i=\frac{1}{2}\text{Tr}\left[T^F  \left\{T^i_a,T^i_b \right\}\right]=0$. Here $T^i_a$ are the generators of the SM groups and $T^F_a$ of $U(1)_{\rm{F}}$ and we have used the normalizations $\left\{Y,Y \right\}=2{\mathcal Y}^2$ and 
$\rm{Tr}(T_a\, T_b)=\frac{1}{2}\, \delta_{ab}$ such that 
$\left\{T_a,T_b \right\}=\frac{1}{2} \, \delta_{ab}$.
Also $A_F=\frac{1}{2} \left[T^{U(1)_Y} \left\{T^F_a,T^F_b \right\}\right]$,  $A^3_F=\frac{1}{2} \left[T^F \left\{T^F_a,T^F_b \right\}\right]$ and $\text{Tr}[T^{U(1)_{F}}]$ must cancel.
For example, we have 
\begin{equation}
 A_1=\frac{1}{2}\text{Tr}\left[T^F \left\{T^{U(1)_Y}_a,T^{U(1)_Y}_b \right\}\right] = \frac{1}{2}\text{Tr}\left[T^F \mathcal{Y}^2 \right] = \frac{1}{2} \sum_\text{ferm.} \left[ \Fmc_\text{ferm.}  \mathcal{Y}^2_\text{ferm.} \right],
\end{equation}
where ``$\text{ferm.}$'' are all the fermions, both those of the SM and the BSM ones.
We write the familiar anomaly cancellation expressions in terms of the
family-dependent charges\footnote{They are written in such a way that
$A_1=5/3\, A_3$. We use this parametrization because it is set up for
gauge coupling unification. An atlas of flavour-dependent $U(1)$ charges
has been given in \cite{Allanach:2018vjg}.},
\bea
 A_1&=&\frac{1}{6}\,\sum^{3}_{i=1} \left[ \Fmc_{\QL i}  -  8\, \Fmc_{u_R\,  i} - 2\, \Fmc_{d_R\, i} + 3\, \Fmc_{\LL \, i } -\, 6    \Fmc_{e_R \,  i} \right] +  X_1, \label{eq:A1}\\
A_2&=&\frac{1}{2}\, \sum^{3}_{i=1} \,  \left[ 3\times  \Fmc_{\QL \, i} +  \Fmc_{\LL\, i}  \right] + X_2, \label{eq:A2}  \\ 
A_3&=&\frac{1}{2}\, \sum^{3}_{i=1} \,  \left[ 2\times \Fmc_{\QL\, i} - \Fmc_{u_R\, i} -  \Fmc_{d_R\, i}  \right] + X_3, \label{eq:A3} \\
A_F&=&\sum^{3}_{i=1}\, \left[  \Fmc_{\QL \, i }^{\ 2} - 2\, \Fmc_{u_R \, i}^{\ 2} +  \Fmc_{d_R\, i }^{\ 2} - \Fmc_{\LL\, i}^{\ 2} + \Fmc_{e_R\, i}^{\ 2}  \right] + X_F,\label{eq:AF} \\
A_F^3 &=& \sum^{3}_{i=1}\, \left[6 \, \Fmc^3_{\QL\, i} - 3 \left( \Fmc^3_{u_R\, i} + \Fmc^3_{d_R\, i}   \right)  +  2\, \Fmc^3_{\LL\, i}   - \Fmc^3_{e_R\, i}\right]\, + \, X_F^3, \label{eq:AF3} 
\eea
Here $X_i$ accounts for the contribution of fermions that are not present in the SM, hence \emph{exotic} fermions, but could be needed for anomaly cancellation. For $A_1$ and $A_F$ color and doublet factors have been included and factorized, for example for $A_F$: 
\bea
& &2\, U(1)_F^2\, U(1)_Y =
\nonumber\\
& & \frac{1}{2} \sum_{i=1}^3 \left[ 3\times 2\times \Fmc_{\QL\, i }^{\ 2} \, \frac{1}{3} + 3 \times \Fmc_{u_R \, i}^{\ 2} \, \frac{-4}{3}  + 3 \times \Fmc_{d_R \, i}^{\ 2} \, \frac{2}{3} + 2 \times \Fmc_{\LL\, i} \, (-1) + \Fmc_{e_R\, i}\, 2 \right]
\nonumber\\
&& {}+ X_F \,.
\eea
If we have
\bea
A_2=A_3=\frac{3}{5} A_1 \,,
\eea
we can achieve gauge coupling unification.

Assuming $N_g$ generations of exotic \emph{leptons}, each containing
$N_D$ left-handed doublets $F_L$ and $N_S$ right-handed singlets $f_R$,
\bea
X_1 &=& \sum_{g=1}^{N_g} \left[
 \sum_{i=1}^{N_D}\, \Y^2_{F_{L_{g,i}}} \left( 2 \, \Fmc_{F_{L_{g,i}}} \right) -
 \sum_{i=1}^{N_S}\, \Y^2_{f_{R_{g,i}}} \left( \Fmc_{f_{R_{g,i}}} \right) \right],
\nonumber\\
X_2 &=& 
\sum_{g=1}^{N_g}\, \sum_{i=1}^{N_D} \left( \Fmc_{F_{L_{g,i}}} \right) ,
\nonumber\\
X_3 &=& 0 \,,
\nonumber\\
X_F &=& \sum_{g=1}^{N_g} \left[
 \sum_{i=1}^{N_D}\, \Y_{F_{L_{g,i}}} \left( 2 \, \Fmc^2_{F_{L_{g,i}}} \right) -
 \sum_{i=1}^{N_S}\, \Y_{f_{R_{g,i}}} \left( \Fmc^2_{f_{R_{g,i}}} \right) \right] ,
\nonumber\\
X_F^3 &=& \sum_{g=1}^{N_g} \left[
 \sum_{i=1}^{N_D} \left( 2 \, \Fmc_{F_{L_{g,i}}}^{3} \right) -
 \sum_{i=1}^{N_S} \left( \Fmc_{f_{R_{g,i}}}^{3} \right) \right] ,
\nonumber\\
X_{\Y} &=& \sum_{g=1}^{N_g} \left[    
 \sum_{i=1}^{N_D}\, 2 \, \Y_{F_{L_{g,i}}} -
 \sum_{i=1}^{N_S}\, \Y_{f_{R_{g,i}}} \right],
\label{eq:AnomCanExtraF}
\eea
where $X_{\Y}$ is the contribution from possible exotic fermions to the cubic $U(1)_Y$ anomaly, which is like \eq{eq:AF3} with the replacement $c_f \rightarrow \mathcal{Y}_f$, and that we will denote by $U(1)^3_Y$.
 Additionally, we have the $U(1)_F$ gauge-gravity anomaly 
\bea \label{eq:AGG}
A_\text{GG} &=& \sum_{i=1}^3
\left[ 6 \, \Fmc_{\QL\, i} - 3 \left( \Fmc_{u_R\, i} + \Fmc_{d_R\, i} \right) + 2\, \Fmc_{\LL\, i} - \Fmc_{e_R\, i} \right] +
X_\text{GG} \,,
\eea
where
\bea \label{eq:XGG}
X_\text{GG} &=& \sum_{g=1}^{N_g} \left[
 \sum_{i=1}^{N_D} \left( 2 \, \Fmc_{F_{L_{g,i}}} \right) -
 \sum_{i=1}^{N_S} \left( \Fmc_{f_{R_{g,i}}} \right) \right] .
\eea

If there are also exotic \emph{quarks}, we obtain
\bea
X_3 &=& \sum_{g=1}^{N_g} \left[
 \sum_{i=1}^{N_{D_q}} \left( 2 \, \Fmc_{F_{L_{g,i}}} \right) -
 \sum_{i=1}^{N_{S_q}} \left( \Fmc_{f_{R_{g,i}}} \right)
\right]
\eea
and have to modify Eqs.~\eqref{eq:AnomCanExtraF} and \eqref{eq:XGG} to
include factors of $3$ for the number of colors.

Potential dark matter candidates are the $U(1)_F$ gauge boson $Z'$ and
the lightest exotic neutral fermion mass eigenstate.  
We will not investigate this aspect in this work, however.

We have to avoid mass mixing between electrically charged SM fermions
and exotic fermions, since this would lead to unacceptably large
tree-level flavour-changing neutral currents.
Hence, the total $U(1)_F$ charge of those combinations
\[
 \overline{F}_{\text{SM}\alpha} H_\beta f_{R_\gamma} \quad,\quad
 \overline{F}_{L_\alpha} H_\beta f_{\text{SM}\gamma} \quad,\quad
 \overline{F}_{\text{SM}\alpha} \phi_\beta F_{L_\gamma}^c \quad,\quad
 \overline{f}_{\text{SM}\alpha} \phi_\beta f_{R_\gamma}^c
\]
that are not forbidden by the SM gauge symmetry has to be non-zero,
where $\alpha$, $\beta$ and $\gamma$ run over all the possible fermions
and scalars, and $F_\text{SM}$, $f_\text{SM}$, represent
respectively SM doublets or singlets.

In order to make the exotics sufficiently heavy, we have to be able to
write down mass terms with masses much larger than the EW scale.  Thus,
the total $U(1)_F$ charge of a sufficient number of combinations
\[
 \overline{F}_{L_\alpha} \phi_\beta F_{L_\gamma}^c \quad,\quad
 \overline{f}_{R_\alpha} \phi_\beta f_{R_\gamma}^c \quad,\quad
 \overline{F}_{L_\alpha} F_{L_\gamma}^c \quad,\quad
 \overline{f}_{R_\alpha} f_{R_\gamma}^c
\]
has to vanish.

The easiest possibility to satisfy the complete set of anomaly
cancellation conditions and further constraints is to introduce two
singlet fermions with opposite hypercharges and $U(1)_F$ charges. This is
because the triangle anomaly $U(1)_Y^3$ is cancelled by the SM matter
content, so every additional matter with non-trivial hypercharge needs
to come in pairs to satisfy the anomaly of $U(1)_Y^3$. This condition
can easily been seen from the last equation of
Eqs.~\eqref{eq:AnomCanExtraF}.

\subsection{A Simple Parameterization to Cancel Anomalies}

In the context of family-dependent $U(1)$ symmetries, generating the
hierarchy of masses through the powers of an expansion parameter inversely
proportional to the Planck scale, Jain and Shrock introduced fermion
mass matrices based on a flavour- and generation-dependent $U(1)$
\cite{Jain:1994hd}. They found that the following parametric sums solve
the equations (\ref{eq:A1}--\ref{eq:AF}):
\bea
&&\sum_{i=1}^3 \, \Fmc_{Q_L \,  i}  =x + v, \quad \sum \Fmc_{d_R\,  i}=-( w + y) , \quad \sum \Fmc_{u_R \, i}=-(2 v + x), \quad \sum \Fmc_{L\, i}= y,\nonumber \\
&& \quad \sum  \Fmc_{e_R\, i}= -x \,. \label{eq:solvparam} 
\eea
In the supersymmetric case, the terms  $X_1$, $X_2$ and $X_F$ in
Eqs.~(\ref{eq:A1}--\ref{eq:A2}) and \eq{eq:AF} respectively  would correspond to
\bea
X_1 &=& \Fmc_{H_u} + \Fmc_{H_d} \,,\nonumber\\
X_2 &=& \Fmc_{H_u} + \Fmc_{H_d} \,,\nonumber\\
X_F &=& \Fmc_{H_d}^2-\Fmc_{H_u}^2 \,,
\eea
where $\Fmc_{H_i} $ are the charges of the Higgsinos and having only one family in that case, we have\footnote{We have rescaled the solutions of \eq{eq:solvparam} and the following with respect to those in \cite{Jain:1994hd} by an overall factor of 3, and hence $z/3 \rightarrow z$.}
\bea
 \Fmc_{H_d} =  (v+w) + z, \quad   \Fmc_{H_u} = - z.
\eea
Taking the parameterizations of \eq{eq:solvparam} only, and not that of the Higgsinos as this would apply only to the supersymmetric case, we have\footnote{In  the supersymmetric case $X_1$, $X_2$ and $X_F$ would be $X_1 =v+w+z$,
$X_2 =-z$,
$X_F = (v+w)^2-2\, z\, (v+w)$ and
$X_3=0$.}
\bea
\label{eq:ACPar}
6\, A_1 &=& 
17 v + 2 w + 5 (3 x + y) +X_1, \nn\\
2\, A_2 &=& 3x +y +3v + X_2, \nn\\
2\, A_3 &=&
3 v +  3 x + y + X_3,\nn\\
2\, A_F &=& -7 v^2 - 6 v x + w (w + 2 y) +X_F.
\eea
What is important to notice is that the
parameterizations of \eq{eq:solvparam} are general, and hence independent
of whether the theory is supersymmetric or not. In particular, it can have different
solutions.
One of these solutions, which fits our needs of having extra singlets
but not doublets or triplets, is to parameterize
\bea
X_1 &=&v+w=0 \ \rightarrow \ v=-w \,.
\eea
Then the rest of the conditions to satisfy Eqs.~(\ref{eq:A1}--\ref{eq:AF}), in the form of \eq{eq:ACPar},
provided $X_2=X_3=0$, can be encoded in the equation
\bea
v= - \frac{1}{3}(3x+y) \,.
\eea
We then need to independently satisfy $X_F=0$.\footnote{Note that $2A_F - X_F$ is reduced to 
$- 2 v (3 (v+x) +  y)$ $= -2 v (-(3 x + y ) + 3x + y)=0$.}
$A_F^3$ also needs to be satisfied independently.
We can rename the variables or make some shifts, but the number of variables remains the same. 
The rest of the anomalies to cancel, which are independent of the above
parameterization, are $X_F$, $X_{\mathcal{Y}}$, and $X_F^3$ given in
\eq{eq:AnomCanExtraF} as well as $X_{GG}$ given in \eq{eq:XGG}.
The simplest and most straightforward non-trivial solution is to
consider two fields with vanishing hypercharge, hence avoiding large
couplings to the Z boson, and opposite $U(1)_F$ charges:
\begin{equation}
X_F=0,\quad X_{\mathcal{Y}}=0,\quad X_F^3=0,\quad X_{GG}=0 \quad
\rightarrow \quad \Fmc_{f_1}=-\Fmc_{f_2} \,.
\end{equation}

\section{Solutions for the Constraints on the Form of Yukawa Matrices}
\label{sec:Yukawas}

The motivation of this work is to have a Higgs field coupling to its own
generation, except for those with the lightest vevs, which could have couplings to each family, such that the Yukawa matrices have the form
\begin{equation}
\label{eq:Hierinit}
Y^d_5, Y^u_6 = \begin{pmatrix}
0 & 0 & 0 \\
0 & 0 & 0 \\
0 & 0 & \mathcal{O}(1)
\end{pmatrix} ,\;
Y^d_3, Y^u_4 = \begin{pmatrix}
0 & 0 &0 \\
0 & \mathcal{O}(1) & \mathcal{O}(1) \\
0 & \mathcal{O}(1) & \mathcal{O}(1)
\end{pmatrix} ,\;
Y^d_1, Y^u_2 = \begin{pmatrix}
\mathcal{O}(1) & \mathcal{O}(1) & \mathcal{O}(1) \\
\mathcal{O}(1) & \mathcal{O}(1) & \mathcal{O}(1) \\
\mathcal{O}(1) & \mathcal{O}(1) & \mathcal{O}(1)
\end{pmatrix} .
\end{equation}
Note that in this case the elements above and below the diagonal have
the same charge combination, that is they are \emph{charge symmetric}.
Hence, one simple way to realise the texture of \eq{eq:Hierinit} is to consider
\begin{equation}
M^d = \frac{v_1}{\sqrt{2}} \begin{pmatrix}
\mathcal{O}(1) & \mathcal{O}(1) & \mathcal{O}(1) \\
\mathcal{O}(1) & 0 & 0 \\
\mathcal{O}(1) & 0 & 0
\end{pmatrix}
+ \frac{v_3}{\sqrt{2}} \begin{pmatrix}
0 & 0 & 0 \\
0 & \mathcal{O}(1) & \mathcal{O}(1) \\
0 & \mathcal{O}(1) & 0
\end{pmatrix}
+ \frac{v_5}{\sqrt{2}} \begin{pmatrix}
0 & 0 & 0 \\
0 & 0 & 0 \\
0 & 0 & \mathcal{O}(1)
\end{pmatrix} ,
\end{equation}
and the same structure for $M^u$ with the changes $M^d \rightarrow M^u$, $v_1 \rightarrow v_2$, $v_3 \rightarrow v_4$ and $v_5 \rightarrow v_6$. However, we find these structures are too restrictive in the sense that cancellation of anomalies requires all the charges to be the same.  This is simply because if all elements in the positions $(i,j)$ for $i\neq j$ are allowed then this forces all charges to be the same, and hence all the families would end up coupling to all the Higgs bosons. 
Another possibility is to have
\begin{equation}
\label{eq:PropMd}
M^d = \frac{v_1}{\sqrt{2}} \begin{pmatrix}
0 & \mathcal{O}(1) & \mathcal{O}(1) \\
0 & 0 & 0 \\
0 & 0 & 0
\end{pmatrix}
+ \frac{v_3}{\sqrt{2}} \begin{pmatrix}
0 & 0 & 0 \\
0 & \mathcal{O}(1) & \mathcal{O}(1) \\
\mathcal{O}(1) & 0 & 0
\end{pmatrix}
+ \frac{v_5}{\sqrt{2}} \begin{pmatrix}
0 & 0 & 0 \\
0 & 0 & 0 \\
0 & \mathcal{O}(1) & \mathcal{O}(1)
\end{pmatrix} ,
\end{equation}
and that could also suit the $u$ sector. However in the particular realization of \cref{sec:specificexample} we choose a diagonal $u$ sector.
The structure in \eq{eq:PropMd} imposes the constraints
\begin{align}
\label{eq:sol1}
\Fmc_{Q_L \, 3} &\neq  \Fmc_{Q_L \, 2}  \neq  \Fmc_{Q_L \, 1} \nn\\
\Fmc_{d_R\, 3 } &= \Fmc_{d_R\, 2} \neq \Fmc_{d_R \, 1}\nn,\\
\Fmc_{u_R\, 3 } &= \Fmc_{u_R\, 2 }\neq \Fmc_{u_R \, 1}\nn,\\
\Fmc_{H_{5}} &\neq \Fmc_{H_{3}} \neq \Fmc_{H_{1}}\nn,\\
\Fmc_{H_{6}} &\neq \Fmc_{H_{4}} \neq \Fmc_{H_{2}},\nn\\
\Fmc_{H_{n}} &\neq -\Fmc_{H_{m}}, \nn\\ 
&  {}_{n \, \in \, \{1,3,5\}},\quad  {}_{m\,  \in\, \{2,4,6\}},
\end{align}
where the last condition is necessary to forbid couplings of the type
\eqref{eq:unwQcoups}.
We note, however, that this condition is necessary but not sufficient,
so we still need to ensure that each term in \eq{eq:unwQcoups} vanishes. 
While we can have exactly the same structure for $M^d$ and $M^u$, we find that is easier to fit the CKM matrix if we assign the mixing only to one sector, so one of the conditions $\Fmc_{d_R\, 3 } = \Fmc_{d_R\, 2}$ or $\Fmc_{u_R\, 3 } = \Fmc_{u_R\, 2 }$ must be lifted.
In this case, we have 14 charges that are parameterized in terms of $x,v,y$, via the equations \eq{eq:solvparam}. 
In this case, we have 
\begin{align}
\label{eq:conditionsh1}
 -\Fmc_{Q_L \, 1,2} + \Fmc_{d_R\, 1}  +\Fmc_{H_{n,m}} &\neq 0 \,,\nn\\
 -\Fmc_{Q_L \, 3} + \Fmc_{d_R\, 1 }  +\Fmc_{H_{3}} &= 0 \,,\nn\\
 -\Fmc_{Q_L \, 1} + \Fmc_{d_R\, 2 }  +\Fmc_{H_{1}} &= 0 \,,\nn\\
  -\Fmc_{Q_L \, 2} + \Fmc_{d_R\, 2 }  +\Fmc_{H_{5}} &= 0 \,,\nn\\
  -\Fmc_{Q_L \, 1} + \Fmc_{d_R\, 3 }  +\Fmc_{H_{1}} &= 0 \,,\nn\\  
  -\Fmc_{Q_L \, 2} + \Fmc_{d_R\, 3 }  +\Fmc_{H_{3}} &= 0 \,,\nn\\  
  -\Fmc_{Q_L \, 3} + \Fmc_{d_R\, 3 }  +\Fmc_{H_{5}} &= 0 \,, 
\end{align}
for $n=1,3,5$ and $m=2,4,6$ such that no coupling to any Higgs is possible for the combinations of the first line in \eq{eq:conditionsh1}.

In the absence of doublet exotics, $F_{L,R}$, beyond the SM and
hypercharged exotics, $f_{L,R}$, the only term leading to mass mixing
between SM fermions and exotics would be 
$\overline{F}_{\text{SM}\alpha} H_\beta f_{R_\gamma}$.
The only possible choice of $F_\text{SM}$ is a lepton doublet with $\mathcal{Y}=-1/2$.
Consequently, the Higgs doublet has to have $\mathcal{Y}=-1/2$ as well,
which implies that its vev is in the upper component.
Thus, this term can only result in unproblematic $\nu_L$--$f_R$ mixing.
As a consequence, it need not be forbidden and thus no any additional
condition on the $U(1)_F$ charges has to be imposed in this case.

A possible exotic mass term could appear from a term like $\overline{F}_{L_\alpha} \phi F_{R_\gamma}$, but since we do not have exotic doublets this is not an issue for our model. 
Hence, it remains to ensure that, for example,
$\overline{f}_{R_\alpha} \phi f_{R_\gamma}^c$ or
$\overline{f}_{R_\alpha} f_{R_\gamma}^c$ is allowed by $U(1)_F$, which results in conditions of the form
\begin{align}
\label{eq:fLphifR_cond}
-\Fmc_{f_{R_{g,i}}} + c_{\phi_n} - \Fmc_{f_{R_{h,j}}} = 0 
\quad\text{or}\quad
\Fmc_{f_{R_{g,i}}} + \Fmc_{f_{R_{h,j}}} = 0 \,.
\end{align}
However, these conditions do not have to be satisfied for \emph{all} possible combinations of $g,h,i,j$. As we have $N_g N_S$ right-handed exotics, a suitable subset of $N_g N_S$ conditions is sufficient.

Given these conditions, specifically the solutions for the charges in \eq{eq:conditionsh1} lead us to finding fifteen different solutions that we specify in Tabs.~\ref{tbl:additional_solutions} and \ref{tbl:additional_solutions_cont}. In the following subsection we present an specific example which corresponds to the eleventh case in \Tabref{tbl:additional_solutions_cont}.

\subsection{Specific Example\label{sec:specificexample}}
In \Tabref{tbl:ModA} we specify the charges for example A\@.
\begin{table}
\centering
\begin{tabular}{|c|cccccll|}
\hline
{\small{Generation/Charges}}& $ \Fmc_{Q_{L\, i}}$  &$\Fmc_{u_{R\, i}}$ &$ \Fmc_{d_{R\, i}}$ & $\Fmc_{L_{L\, i}}$ & $\Fmc_{e_{R\, i}}$ & $\Fmc_{H_{n}}$ & $\Fmc_{H_{m}}$ \\
\hline
{\tiny{$i=1$}} & 7  & $0$ & $-2$ & $-10$ & $-6$ & $\Fmc_{H_1}= 5$ & $\Fmc_{H_2}= 7$ \\
{\tiny{$i=2$}} & $-2$  & $-1$ & $2$  & 4  &  4  & $\Fmc_{H_3}=-4$ & $\Fmc_{H_4}=-1$ \\
{\tiny{$i=3$}} & $-6$ & $-3$ &  2  & 9  &  8 & $\Fmc_{H_5}=-8$ & $\Fmc_{H_6}= -3$ \\
\hline
{\small{New fermions}}& $\Fmc_{\chi_1}$ & $\Fmc_{\chi_2}$  &  &   &  &  &\\
\hline
 & $-1$ & $1$ & & & & & \\
\hline
{\small{Scalars breaking $U(1)_F$}} & $\Fmc_{\phi_1}$ & $\Fmc_{\phi_2}$ & $\Fmc_{\phi_3}$ & & & & \\
\hline
 & $-2$ & $7$ & $3$ & & & & \\
\hline
\end{tabular}
\caption{Family symmetry charges for model $U(1)_F$ A\@.
 All new fermions and scalar singlets have vanishing hypercharge.}
\label{tbl:ModA}
\end{table}
They allow the form of the matrices in \eq{eq:PropMd} for the $d$ sector while keeping the $u$ sector diagonal. 
This is only one example of charges. We have found $15$ different anomaly-free
solutions that satisfy the conditions of \eq{eq:conditionsh1}.
Example A contains one family of right-handed exotics with
opposite charges, allowing them to become sufficiently heavy by means of
the gauge-invariant mass term
\begin{equation} \label{eq:ExoticMass}
\mathcal{L}_\text{mass}^\chi =
\overline{\chi_1} M_\chi \chi_2^c + \text{h.c.}
\end{equation}
Additional contributions to the exotics masses come from coupling to one
of the scalar singlets,
\begin{equation} \label{eq:ExoticYukawa}
\mathcal{L}_\text{Yuk}^\chi =
\overline{\chi_1} \phi_1 \chi_1^c + 
\overline{\chi_2} \phi_1^\dagger \chi_2^c + 
\text{h.c.}
\end{equation}
The charge assignment also cancels the gravitational anomaly
$X_{GG}=\Fmc_{\chi_1}+\Fmc_{\chi_2}$ and cubic anomaly
$X_F^3=\Fmc_{\chi_1}^3+\Fmc_{\chi_2}^3$. All the other anomalies cancel
automatically since the exotics have zero hypercharge.

We do not impose additional conditions to obtain particular forms of the
lepton mass matrices because the purpose of this work is to give a proof
of principle that there exist solutions with the desired hierarchy for
the \emph{quark} sector.  We only verify that the $U(1)_F$ charge
assignment allows for a sufficient number of non-vanishing mass terms.
For the charged leptons, the allowed combinations in example A are
\[
\overline{L}_{L3} H_1 e_{R2} \quad,\quad
\overline{L}_{L1} H_3 e_{R1} \quad,\quad
\overline{L}_{L2} H_3 e_{R3} \quad,\quad
\overline{L}_{L3} \, \epsilon H_4^* e_3 \,,
\]
i.e., the charged lepton mass matrix has the form
\begin{equation} \label{eq:Me}
M^e= \frac{1}{\sqrt{2}}\left(\begin{array}{ccc}
Y^e_{11} v_3 & 0 & 0 \\
0 & 0 & Y^e_{23} v_3 \\
0 & Y^e_{32} v_1 &  Y^e_{33} v_4
\end{array}
\right),
\end{equation}
which can give the right eigenvalues.
Indeed they can be easily computed to yield
\bea
m_e&=&\frac{v_3}{\sqrt{2}} Y^e_{11} \,,
\nonumber\\
m_\tau,m_\mu&=&\frac{v_4}{2\sqrt{2}} Y^e_{33}\pm
\frac{\sqrt{4 v_1 v_3 Y^e_{23} Y^e_{32} + v_4^2 (Y^e_{33})^2}}{2\sqrt{2}} \,,
\eea
which shows that it is possible to fit the charged lepton masses.

For generality, we assume that some heavy fields lead to dimension-$5$ operators
of the form
\[
	\left( \overline{L_L} \epsilon H_l^* \right) \kappa
	\left( H_{l'}^\dagger \epsilon L_L^c \right)
	\quad,\quad
	\left( \overline{L_L} H_k \right) \kappa'
	\left( H_{k'}^T L_L^c \right)
	\quad,\quad
	\left( \overline{L_L} H_k \right) \kappa''
	\left( H_l^\dagger \epsilon L_L^c \right)
\]
with
$l,l' \in \{1,3,5\}$ and $k,k' \in \{2,4,6\}$,
which respect the $U(1)_F$ symmetry.%
\footnote{This is expected for operators generated by the exchange of scalar $SU(2)_L$ triplets or gauge-singlet fermions. For fermions charged under $U(1)_F$, the discussion becomes model-dependent and more complicated.}
The couplings $\kappa$, $\kappa'$, and $\kappa''$ are symmetric matrices in flavour space.
These operators yield Majorana neutrino masses after EW symmetry breaking.

Given the charge assignments of model A, the resulting neutrino mass
matrix in the gauge eigenstate basis has the form
\begin{equation}
    M^\nu = \begin{pmatrix}
        0 & a & b \\
        a & d & 0 \\
        b & 0 & 0
    \end{pmatrix} .
\end{equation}
According to \eq{eq:Me}, changing to the charged lepton mass eigenstate basis requires a rotation in the 2-3 plane. Denoting the corresponding rotation angle of the lepton doublets by $\theta$, this changes the neutrino mass matrix to
\begin{equation}
    \tilde M^\nu = \begin{pmatrix}
        0 & a \cos\theta - b \sin\theta & a \sin\theta + b \cos\theta \\
        a \cos\theta - b \sin\theta & d \cos^2\theta & d \sin\theta \cos\theta \\
        a \sin\theta + b \cos\theta & d \sin\theta \cos\theta & d \sin^2\theta
    \end{pmatrix} .
\end{equation}
This matrix is subject to the renormalization group evolution from the symmetry-breaking scale to low energy. The changes are expected to be sizable due to the $\mathcal{O}(1)$ Yukawa couplings. As a consequence, determining whether the resulting neutrino mass parameters can be compatible with observations requires a dedicated study beyond the scope of this paper.

For the quark sector, we work out the specific form of the mass matrices.
Using the definition \eqref{eq:HiggsDoublets}, we write the quark part
of the Yukawa Lagrangian \eqref{eq:MatterLagrangian} as
\begin{equation} \label{eq:LQuarkHiggs}
	\mathcal{L}_\text{Yuk} =
	-\!\sum_{l=1,3,5} H_l^0 \, \overline{d_L'} \, Y^d_l \, d_R' \,
	-\!\sum_{k=2,4,6} H_k^0 \, \overline{u_L'} \, Y^u_k \, u_R'
	+\text{h.c.} \,,
\end{equation}
where the primes on the quark fields (understood to be vectors in
flavour space here) denote interaction eigenstates.
Once the scalar doublets have received vevs%
\footnote{We assume a CP-conserving scalar potential for simplicity,
implying real vevs.}
$v_n \in \mathbb{R}$ as specified in \eq{eq:H0Decomposition}, we can
write
\begin{align}
	\mathcal{L}_\text{mass} &=
	-\overline{d'_L} M^d d'_R - \overline{u'_L} M^u u'_R + \text{h.c.} \,,
\nonumber\\
	M^d &= \frac{1}{\sqrt{2}}\sum_{l=1,3,5} v_l Y_l^d \quad,\quad
	M^u = -\frac{1}{\sqrt{2}}\sum_{k=2,4,6} v_k Y_k^u \,,
 \label{eq:Mquarks}
\end{align}
where
\begin{align}
	\sum_{n=1}^6 \frac{v_n^2}{2} &= \left( 175\GeV \right)^2 .
\end{align}
Note the sign in the expression for $M^u$ in \eq{eq:Mquarks}, which
comes from our definition of $H_k^0$ in \eq{eq:H0Decomposition}.
Following the conditions \eqref{eq:conditionsh1} the form of the
down-type quark mass matrix becomes
\begin{equation}
\label{eq:Mdform}
M^d=\frac{1}{\sqrt{2}}\left(
\begin{array}{ccc}
0  & v_1 (Y^d_1)_{12}     & v_1 (Y^d_1)_{13}\\
0 &   v_3 (Y^d_3)_{22} & v_3 (Y^d_3)_{23}\\
v_3 (Y^d_3)_{31} &   v_5 (Y^d_5)_{32} &   v_5 (Y^d_5)_{33}\\
\end{array}
\right),
\end{equation}
and the up-type mass matrix has a diagonal form,
\begin{equation}
\label{eq:formofMd_MA}
M^u=-\frac{1}{\sqrt{2}}\left(
\begin{array}{ccc}
v_2 (Y^u_2)_{11} & 0 &0 \\
0  &  v_4 (Y^u_4)_{22}& 0\\
0  &  0 & v_6 (Y^u_6)_{33}
\end{array}
\right).
\end{equation}
Note that for $M^u_i>0$ we need to set $(Y^u)_{ii}<0$, since $v_{2,4,6}>0$.
For the diagonalisation of the mass matrices we use the convention
\begin{equation} \label{eq:diagconvmat}
\begin{split}
\widehat{M^q} & = V^q_L \, M^q\, V^{q \dagger}_R \;, \quad q=u,d \,,
\\
V^q_L\, q'_L &= q_L \quad,\quad V^q_R\, q'_R= q_R \,,
\end{split}
\end{equation}
where $q_L$ and $q_R$ are mass eigenstates and $\widehat{M^q}$ is diagonal and real. In this convention,
$V_\ckm=V^u_L V^{d\dagger}_L$. In this way, it is easy to fit numerical
values of the matrices, since the square of the down-sector mass matrix is given only in terms of the mass values and the CKM matrix,
\begin{equation}
\label{eq:Mdsq}
M^dM^{d\dagger}=V_\ckm \left( \widehat{M^d} {\widehat{M^d}}^\dagger \right) V_\ckm^\dagger \,.
\end{equation}
In the appendix we give the explicit numerical values of the matrix $M^d$ obtained in terms of the matrix of \eq{eq:Mdsq} above.

\section{Scalar Potential and Phenomenology} 
\label{sec:scalarpotential}

\subsection{Scalar Potential}
A complete analysis of the scalar potential is beyond the scope of this
paper, so we restrict ourselves to a general discussion and present a
benchmark point that may be phenomenologically viable.  First of all,
the general potential with only one singlet can be written as
\begin{align*}
V &= \mu_\phi^2 \, \phi^\dagger\phi + \mu_{nr}^2 H_n^\dagger H_r +
\left( \tilde\mu_{nr}^2 H_n^T \epsilon H_r + \text{h.c.} \right)
\nonumber\\
{}&+
\left[ \mu_{\phi\,nr} \, \phi H_n^\dagger H_r +
\left( \tilde\mu_{\phi\,nr} \, \phi + \tilde\mu'_{\phi\,nr} \, \phi^\dagger \right)
  \left( H_n^T \epsilon H_r \right) +
\text{h.c.} \right]
\nonumber\\
{}&+
b_\phi \left(\phi^\dagger\phi\right)^2 +
b_{nrms} \left(H_n^\dagger H_r\right) \left( H_m^\dagger H_s \right) +
b_{\phi\,nr} \left(\phi^\dagger\phi\right) \left(H_n^\dagger H_r\right)
\nonumber\\
{}&+
\left[
 b'_{\phi\,nr} \phi\phi \left(H_n^\dagger H_r\right) +
 \left( \tilde b'_{\phi\,nr} \phi\phi + \tilde b''_{\phi\,nr}
  \phi^\dagger \phi^\dagger \right) \left(H_n^T \epsilon H_r\right) +
 \tilde b_{\phi\,nr} \left(\phi^\dagger\phi\right) \left(H_n^T \epsilon H_r\right) +
 \text{h.c.}
\right]
\nonumber\\
{}&+
\left[
 \tilde b_{nrms} \left(H_n^\dagger H_r\right) \left( H_m^T \epsilon H_s \right) +
 b'_{nrms} \left(H_n^T \epsilon H_r\right) \left( H_m^T \epsilon H_s \right) +
 \text{h.c.}
\right]
,
\end{align*}
where of course some couplings are forbidden by the gauge symmetry.
This potential is not phenomenologically viable since it contains
several accidental symmetries that are spontaneously broken, leading to
the appearance of massless pseudoscalars.  Hence, we are forced to
introduce several singlets and assign charges such that no
accidental global symmetries appear.  
The model $U(1)_F$ A presented in \Tabref{tbl:ModA}
satisfies this requirement.  For this model, the potential is
\begin{align}
V &= \sum_{p=1}^3\mu_{\phi_p}^2 \phi^\dagger_p\phi_p +
\sum_{n=1}^6 \mu_{nn}^2 H_n^\dagger H_n +
\left[ \mu_{\phi_1\,64} \, \phi_1 \, H_6^\dagger H_4 + \text{h.c.} \right]
\nonumber\\
&+ \left[
\tilde\mu_{\phi_1\,16} \, \phi_1 \left( H_1^T \epsilon H_6 \right) +
\tilde\mu_{\phi_2\,36} \, \phi_2 \left( H_3^T \epsilon H_6 \right) +
\tilde\mu_{\phi_3\,23}'\,\phi_3^\dagger \left( H_2^T \epsilon H_3 \right) +
\text{h.c.} \right]
\nonumber\\
&+
\sum_{p=1}^3 \sum_{q=p}^3 b_{\phi_p\phi_q}
 \left(\phi_p^\dagger\phi_p\right) \left(\phi_q^\dagger\phi_q\right) +
\left[ b_{\phi_1\phi_1\phi_2\phi_3} \, \phi_1^2 \phi_2 \phi_3^\dagger +
 \text{h.c.} \right]
\nonumber\\
&+
\sum_{n=1}^6 \sum_{m=n}^6 b_{nnmm} \left(H_n^\dagger H_n \right)
\left(H_m^\dagger H_m \right) +
\sum_{n=1}^5 \sum_{m=n+1}^6 b_{nmmn} \left(H_n^\dagger H_m \right)
\left(H_m^\dagger H_n \right)
\nonumber\\
&+
\sum_{n=1}^6 \sum_{p=1}^3 b_{\phi_p\,nn} \, \phi^\dagger_p\phi_p \left(H_n^\dagger H_n \right)
\nonumber\\
&+ \left[
b_{\phi_1\phi_2\,13} \, \phi_1^\dagger\phi_2 \, \big(H_1^\dagger H_3\big) +
b'_{\phi_1\phi_1\,53} \, \phi_1\phi_1 \, \big(H_5^\dagger H_3\big) +
b'_{\phi_2\phi_3\,26} \, \phi_2\phi_3 \, \big(H_2^\dagger H_6\big) +
\text{h.c.} \right]
\nonumber\\
&+ \left[
\tilde b_{\phi_1\phi_1\,14}'\,\phi_1\phi_1 \left(H_1^T \epsilon H_4\right)+
\tilde b_{\phi_1\phi_2\,34}'\,\phi_1\phi_2 \left(H_3^T \epsilon H_4\right)+
\tilde b_{\phi_1\phi_3\,25}'\,\phi_1\phi_3 \left(H_2^T \epsilon H_5\right)+
\text{h.c.} \right]
\nonumber\\
&+ \left[
\tilde b_{\phi_1\phi_2\,45} \,\phi_1^\dagger\phi_2 \left(H_4^T \epsilon H_5\right)+
\text{h.c.} \right]
\nonumber\\
&+ \left[
\widehat{\tilde{b}}_{3514}\, (H_3^\dagger H_5)(H_1^T \epsilon H_4) + 
\widehat{\tilde{b}}_{3154}\, (H_3^\dagger H_1)(H_5^T \epsilon H_4) +
\widehat{\tilde{b}}_{4616}\, (H_4^\dagger H_6)(H_1^T \epsilon H_6) +
\text{h.c.} \right]
\nonumber\\
&+ \left[
\widehat{\tilde{b}}_{3415}\, (H_3^\dagger H_4)(H_1^T \epsilon H_5) +
\text{h.c.} \right] .
\label{eq:PotentialC}
\end{align}
All of the possible terms of the type $\tilde{b}_{nrms}$ with the
possible interchanging of sub\-indices $n,r,m$ and $s$ have been grouped
in the terms $\widehat{\tilde{b}}_{nrms}$.
Expanding the terms with couplings
$\widehat{\tilde{b}}_{3514}$ and $\widehat{\tilde{b}}_{3154}$ into their
components, we see that they differ only in terms containing two charged
and two neutral scalars; consequently, only the sum of the couplings
appears in the neutral scalar and pseudoscalar masses, and we could omit
one of them when analyzing these masses.
Besides, the expansion of the term with coupling
$\widehat{\tilde{b}}_{3415}$ contains only contributions with charged
scalars; thus, this term is irrelevant for neutral scalar masses as well.

After symmetry breaking, scalars and pseudoscalars mix in general.
Assuming CP conservation in the scalar potential for simplicity,
we arrive at the mass eigenstates
\begin{align}
	h^s_m &= \left(S_\sigma\right)_{mn} \sigma_n \quad,\quad
	h^p_m = \left(S_\varphi\right)_{mn} \varphi_n
	\quad,\quad m,n = 1,\dots,9 \,,
\nonumber\\
	h^+_m &= {(S_+)}_{ma} H^+_a \quad,\quad m, a=1,\dots,6 \,,
\label{eq:ScalarMixing}
\end{align}
where $\sigma_n$, $\varphi_n$ and $H^+_a$ are the interaction eigenstates,
and $S_\sigma$, $S_\varphi$ and $S_+$ are orthogonal matrices.
The decomposition of the neutral components of the Higgs doublets into
real fields is given in \eq{eq:H0Decomposition}.  For the SM singlets
breaking $U(1)_F$, we use
\begin{equation} \label{eq:PhiDecomposition} 
	\phi_1 = \frac{1}{\sqrt{2}} \left( v_7 + \sigma_7 + i \varphi_7 \right)
\end{equation}
and analogously for $\phi_2$ and $\phi_3$.

\subsection{Phenomenology} \label{sec:HiggsPheno}
In this section we discuss constraints related to the scalar mass spectrum and possible observational implications.
The study of constraints from flavour-changing neutral currents and CP
violation is left for \cref{sec:fcnc}.
Obviously, the scalar potential has a huge number of free parameters and
dedicated analyses would be required to study its phenomenology thoroughly.
As this is not our purpose (and also as model $U(1)_F$ A is unlikely to
be the most elegant realization of the scenario), we restrict ourselves
to present one acceptable benchmark point as a proof of principle.  That
is, we set all parameters to particular values resulting in a scalar
sector that is not in conflict with observations.  These values
are given in \cref{tab:coeficientsb}.  The lightest scalar $h^s_1$ with a
mass of about $125\GeV$ is composed predominantly of the doublet component
$\sigma_6$ coupling to the top quark.  The admixtures of the other
doublet components $\sigma_1,\dots,\sigma_5$ have the correct values to
ensure SM-like couplings of $h^s_1$ to all SM fermions.  All additional
scalars, including the charged ones, and most of the pseudoscalars have
masses in the multi-TeV range and are thus unaffected by bounds from
current collider searches.
However, the masses of the third-lightest scalar and the second-lightest 
pseudoscalar, which are mainly composed of the doublet $H_5$, are around $15\:$TeV.  
Hence, these particles can be produced at a next-generation collider such as the Future Circular Collider (FCC) with center of mass energy $100\:$TeV.

There is one pseudoscalar, $h^p_3 \equiv a$,%
\footnote{The pseudoscalars $h^p_1$ and $h^p_2$ are the would-be
Nambu-Goldstone bosons that are eaten by the $Z$ and the $Z'$,
respectively.}
whose mass cannot be raised above the EW scale, which is 
one of the unique features of our model.  At the presented
benchmark point, it is about $15\GeV$.
When $b_{\phi_1\phi_1\phi_2\phi_3} \to 0$, the scalar potential gets an additional $U(1)$ symmetry which can be identified with the Peccei-Quinn symmetry~\cite{Ellwanger:2009dp}. The light pseudoscalar becomes a massless Goldstone boson, so the small mass is technically natural.
However, this state is mainly composed of the singlet components
$\varphi_8$ and $\varphi_9$.  Its admixture of $\varphi_6$ is of order
$10^{-6}$, which suppresses the coupling to the top quark.  The largest
admixture of a doublet component is of order $10^{-3}$, which ensures
highly suppressed couplings to the EW gauge bosons as well.

Since the coupling of the light pseudoscalar to the top quark is suppressed, the cross section for producing this particle at the LHC is much
smaller than the cross section for producing the observed Higgs boson
and thus compatible with experimental bounds.
The contribution to invisible Higgs decays is sufficiently suppressed by
choosing small values for the relevant couplings $b_{\phi_2\,66}$ and
$b_{\phi_3\,66}$.  These coefficients have no impact on the masses
of the lightest scalar and pseudoscalar.  For the example we present, we
fix both of them to $10^{-5}$.

The light pseudoscalar can be searched for in decays of the SM-like
lightest scalar $h^s_1$. Its main decay channel is $a \to b \bar{b}$. The relevant interaction is 
\begin{equation}
\mathcal{L} \supset
-\frac{1}{2} \mu_{haa} \, h^s_1 a a +
\frac{i}{\sqrt{2}} w_{abb} \, a\, \bar{b} \gamma_5 b
\end{equation}
with $\mu_{haa}\simeq -1.5 \GeV$ and $w_{abb}=0.0027$ for the benchmark
point. The predicted branching fraction is
$\mathcal{B}(h^s_1 \to aa) \simeq 0.038$ setting the total width of $h^s_1$ equal to the SM Higgs width of $4.6\MeV$.

\begin{figure}
\centering
$\vcenter{\hbox{\includegraphics[width=0.45\linewidth]{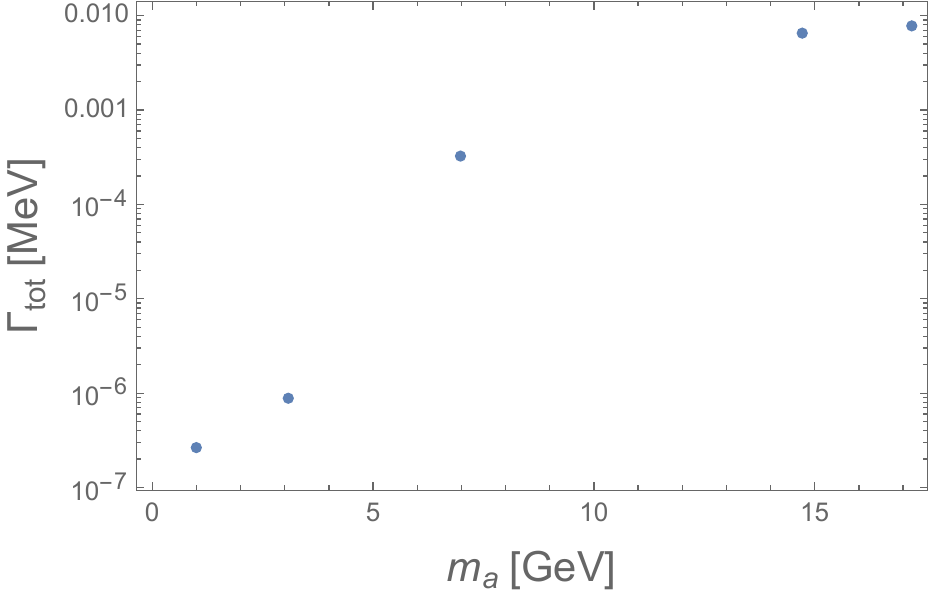}\qquad\includegraphics[width=0.45\linewidth]{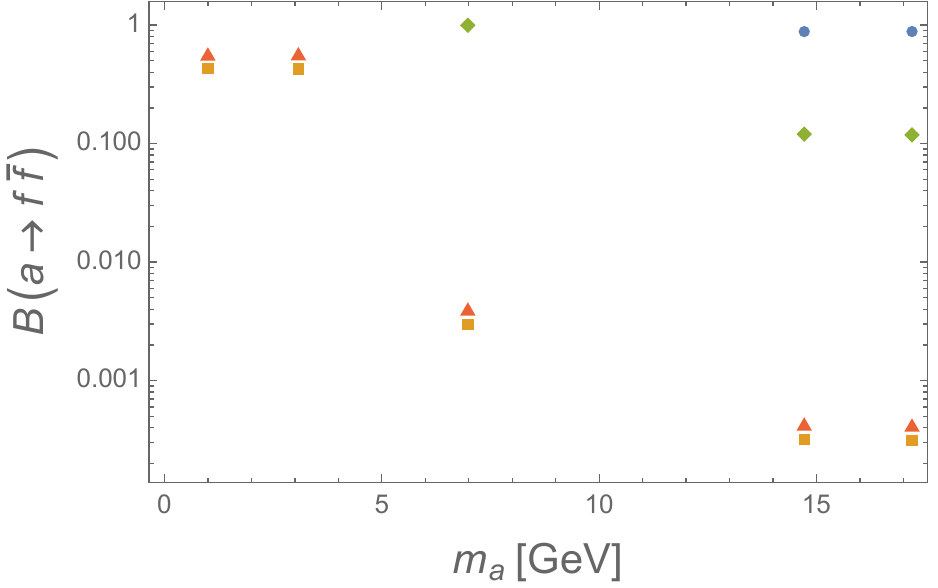}}}$
\caption{The total decay width of the lightest pseudoscalar $a$, $\Gamma_\text{tot}$ (left panel), and the branching fractions of the main decay channels of $a$, $\mathcal{B}(a \to f \bar{f})$ with $f=b, \tau, \mu, s$ (circles, diamonds, triangles, and squares, respectively; right panel). Both are plotted as a function of $m_a$.}
\label{fig:ma_decay}
\end{figure}

We obtain the total decay width of $a$ to be $\Gamma_a = 6.5 \times 10^{-3} \MeV$, showing that it decays promptly inside the detector.  The branching fractions are $\mathcal{B}(a\to f\bar{f})=0.88,0.00032,0.12,0.00043$ for $f=b,s,\tau,\mu$, respectively. The decays of $a$ into other channels are negligible.
The current experimental bounds on $\sigma_{h^s_1}/\sigma_h^\text{SM}\mathcal{B}(h^s_1 \to aa \to fff'f')$ are collected in~\cite{Cepeda:2021rql}.
Assuming the production cross section of the lightest scalar to be the
same as that of the SM Higgs, our model prediction is about one order of magnitude below the current bound for $h^s_1 \to aa \to bb\tau\tau$ and further below the bounds for the other channels.
The HL-LHC, where the luminosity is increased by a factor 10 compared to the LHC's design value, may be promising for detecting the light pseudoscalar in the $bb\tau\tau$ channel. 
In Fig.~\ref{fig:ma_decay} we show the total decay width of $a$ and the branching fractions of the main decay channels as a function of $m_a$ for five benchmark masses.

We do not have to worry about $Z$ decays into the light pseudoscalar
either.  First, a spin-1 particle cannot decay into a pair of identical
scalar bosons because of total angular momentum conservation and the
spin-statistics theorem.  If $Z\rightarrow aa$ happens,
$J(Z)=J(aa)=L(aa)=1$ since $a$ has no spin. Then the wavefunction of the
$aa$ system will have the factor $(-1)^2 \times (-1)^{L{(aa)}}=-1$ under
the exchange of two identical $a$'s, which contradicts the
spin-statistics theorem. This is analogous to the Landau-Yang theorem,
and it is the reason why $\rho^0 (770)$ decays into $\pi^+ \pi^-$ and not
into $\pi^0 \pi^0$ even though it is kinematically allowed.
Second, a similar argument holds for $Z\rightarrow aaa$.  For example,
$Z\rightarrow aaa$ will be described by
\[
Z_\mu \partial^\mu a^3 = - ( \partial^\mu Z_\mu ) a^3 = 0
\quad,\quad
Z_{\mu\nu} (\partial^\mu) (a \partial^\nu a) a = 0 \,, \text{ etc.}
\]
in terms of the Lagrangian.
Third, $Z^0 \to a \gamma$ occurs only at the one-loop level.
Thus, all $Z$ decays involving the light pseudoscalar are strongly
suppressed not only by the small coupling but also by the fact that they
have final states with at least three particles or proceed via loops.

For pseudoscalar masses below the $\Upsilon$ mass ($9.460 \GeV$), 
the decay $\Upsilon \to \gamma a$ becomes allowed in principle, but it is 
strongly suppressed as well.
Its decay rate normalized to $\Upsilon \to \mu^+ \mu^-$ is given by 
(ignoring QCD corrections to both decays) \cite{Wilczek:1977zn,Vysotsky:1980cz,Nason:1986tr}
\begin{equation}
\frac{\mathcal{B} ( \Upsilon(1S)  \to \gamma a)}{\mathcal{B} ( \Upsilon(1S) \to \mu^+ \mu^- )} 
=
\frac{ |w_{abb}|^2}{4 \pi \alpha} \left( 1 - \frac{m_a^2}{m_\Upsilon^2} \right) ,
\end{equation}
so
\begin{equation}
\mathcal{B}( \Upsilon(1S) \to \gamma a) \lesssim 2 \times 10^{-6}
\end{equation}
for $| w_{abb} | \simeq 3 \times 10^{-3}$ and $\mathcal{B} (
\Upsilon(1S) \to \mu^+ \mu^- ) \simeq 2.48 \%$~\cite{Workman:2022ynf}. This is compatible with the current 
upper limit~\cite{Workman:2022ynf}
\[\mathcal{B}(\Upsilon(1S) \to \gamma a \to \gamma \mu^+ \mu^-)<9 \times 10^{-6} \,, 
\]
but could be reachable in the 
future at high-luminosity $B$ factories such as Belle II.
Constraints from flavour observables will be discussed in the next section.

We have checked numerically that the potential grows for
large field values and is thus bounded from below.  The only remaining
constraint, which would require a dedicated analysis, is the absence of
charge-breaking minima.

A striking signature of the model could be flavour-violating Higgs
decays or Higgs-mediated flavour-changing neutral currents, but they are strongly suppressed. To see this we refer to
\Figref{fig:FeynmanRules} and
Eqs.~(\ref{eq:ScalaLagMassE}--\ref{eq:scalarcoupsME}) in the next
section.  The rotation to the mass eigenstate basis of scalars and
quarks leaves only very small off-digonal terms in the coupling matrix
of the lightest scalar.  In particular, the decay $h^s_1 \to b s$, which
appears to be the most likely one according to the form of the mass
matrix $M^d$ in \eq{eq:Mdform}, is governed by
$\left(\mathcal{V}_d\right)^1_{bs}$ and
$\left(\mathcal{V}_d\right)^1_{sb}$ from \eq{eq:scalarcoupsME}.  For the
example presented here, taking into account the parameter values of
Tabs.~\ref{tab:numericalMd} and \ref{tab:coeficientsb}, these couplings
are suppressed with respect to $\left(\mathcal{V}_d\right)^1_{bb}$ by at
least $5$ orders of magnitude.  Therefore, the decay is unlikely to be
observable by future Higgs factories \cite{Kamenik:2023ytu}.

\subsection{Additional Benchmark Points}

The goal of this work is to present an example of our framework, but in
order to study how the phenomenology could be altered we present four different benchmark points that differ in the spectra of scalars and/or pseudoscalars. We present these benchmark points in Appendix \ref{app:benchmarkpoints}.
From \Figref{fig:ma_decay} we can see that the dominant decay of
$a \to f \bar{f}$ above the $m_b$ threshold is the decay to $f=b$, while the total decay width increases with the increase of the mass $m_a$.

We note that this kind of models has the potential to undergo first order phase transitions. In fact, a potential for a scalar field, $\phi$ that has the form $V(\phi)=m^2 \phi^2 + E \phi^3 + \lambda \phi^4$, with the appropriate temperature corrections (see for example \cite{Linde:1990flp}), can lead to phase transitions depending on the importance of the cubic term. We find that for some directions, the potential exhibits this behavior. For example, for benchmark point 5 there is a scalar around $10^5\GeV$ (see \Tabref{tb:BP5pars}). For temperatures corresponding to this scale this can lead to gravitational waves peaking around a frequency equal to $0.1$\:Hz, and therefore accessible to LISA \cite{LISACosmologyWorkingGroup:2022jok}. Given the intensity of research in this direction, once a realistic model for all fermion sectors is achieved, a study of gravitational waves in this context will be worthwhile.

\section{Flavour-Changing Neutral Currents and CP Violation}
\label{sec:fcnc}

The multi-Higgs scenario induces flavour-changing neutral currents (FCNC)
and CP violation at tree level, which can lead to stringent constraints.
We consider bounds obtained from indirect CP violation in the neutral
kaon system ($\epsilon_K$) and from the mass differences of neutral $B$
mesons ($\Delta m_{B_q}$ for $B^0_q$, where $q=d,s$).
These observables follow from the matrix elements
$\braket{K^0 | H^{ds}_{\Delta S=2} | \overline{K^0}}$ and
$\braket{B_q^0 | H^{qb}_{\Delta B=2} | \overline{B_q^0}}$, respectively,
where $H^{ds}_{\Delta S=2}$ and $H^{qb}_{\Delta B=2}$ are the relevant
effective Hamiltonians.

\paragraph{Vector contributions}
Kinetic and/or mass mixing between $Z$ and $Z'$ does lead to FCNC via $Z$ exchange,
but this will be a subdominant contribution as long as they are small.  At the one-loop level, the usual SM FCNC are
generated by $W$ exchange.

\begin{figure}
\centering
$\vcenter{\hbox{\includegraphics{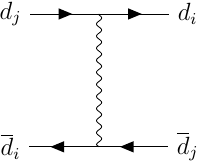} \qquad
\includegraphics{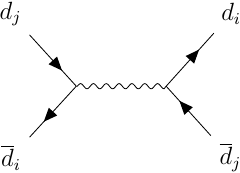} \qquad}}$
\caption{Tree-level contributions to FCNC observables from $Z'$
exchange.  The quark assignments for the meson systems we consider are
given in \cref{tab:Pij}.}
\label{fig:DeltamKVector}
\end{figure}

$Z'$ exchange (\cref{fig:DeltamKVector}) leads to
FCNC due to the generation-dependent $U(1)_F$ charges.
Schematically, the amplitude is of the form $\frac{|g_{sd}|^2}{m^2}$.
This yields the dominant vector contribution unless
the $Z'$ mass is much larger than $m_Z$.

The interactions between $Z'$ and the SM quarks are given by \cite{Ko:2011di}
\begin{align}
\mathcal{L}  &\supset
- g_F Z'_\mu \left[ \overline{u'_L} Q_L^{u} \gamma^\mu u'_L  
+ \overline{u'_R} Q_R^{u}  \gamma^\mu u'_R 
+ \overline{d'_L} Q_L^{d}  \gamma^\mu d'_L 
+ \overline{d'_R} Q_R^{d}  \gamma^\mu d'_R \right] 
\nonumber\\
&= - g_F  Z'_\mu \left[ 
\overline{u_L} \left( V_L^u Q_L^{u} V_L^{u \dagger} \right) \gamma^\mu u_L  
+ \overline{u_R} \left( V_R^u Q_R^{u} V_R^{u \dagger} \right) \gamma^\mu u_R \right.
\nonumber\\
& \hphantom{= - g_F  Z'_\mu} \left. {}+ 
\overline{d_L} \left( V_L^d Q_L^{d} V_L^{d \dagger} \right) \gamma^\mu d_L 
+ \overline{d_R} \left(V_R^d Q_R^{d} V_R^{d \dagger} \right) \gamma^\mu d_R
\right] 
\nonumber\\
&\equiv - g_F Z'_\mu
\left[ (g_L^u )_{ij} \overline{u_{L i}} \gamma^\mu u_{L j}   
+ (g_R^u)_{ij} \overline{u_{R i}} \gamma^\mu u_{R j} 
+ (g_L^d)_{ij} \overline{d_{L i}} \gamma^\mu d_{L j} 
+ (g_R^d)_{ij} \overline{d_{R i}} \gamma^\mu d_{R j} \right] ,
\end{align}
where $u^{\prime\, T}_{L (R)} \equiv ( u' , c' , t' )_{L (R)}$, 
$d^{\prime\, T}_{L(R)} \equiv (d' , s' , b' )_{L(R)}$ are fields in the interaction 
basis.  Similarly for the fields in the mass basis, we have  
$u^T_{L(R)} = (u,c,t)_{L(R)}$, $d^T_{L(R)} = (d,s,b)_{L(R)}$.

The $Z'$ charge matrices in the interaction basis are given by 
\begin{align}
& Q_L^u = Q_L^d = {\rm diag} ( c_{Q_L,1} , c_{Q_L,2}, c_{Q_L,3} )
\quad,\quad
Q_R^u = {\rm diag} ( c_{u_R,1} , c_{u_R,2} , c_{u_R,3} ) \,,
\nonumber\\
&Q_R^d = {\rm diag} ( c_{d_R,1} , c_{d_R,2} , c_{d_R,3} ) \,,
\end{align}
and those in the mass basis are given by 
\[
g_L^u \equiv V_L^u Q_L^{u} V_L^{u \dagger} \quad,\quad
g_R^u \equiv V_R^u Q_R^{u} V_R^{u \dagger} \quad,\quad
g_L^d \equiv V_L^d Q_L^{d} V_L^{d \dagger} \quad,\quad
g_R^d \equiv V_R^d Q_R^{d} V_R^{d \dagger} \,,
\]
respectively.

\begin{figure}
\centering
$\displaystyle
\vcenter{\hbox{\includegraphics[scale=0.75]{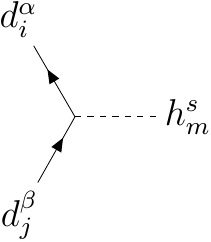}}}
\quad
\frac{-i}{\sqrt{2}} \left( \left(\mathcal{V}_d\right)^m_{ij} P_L +
 \left(\mathcal{V}_d^*\right)^m_{ji} P_R \right) \delta_{\alpha\beta}
$
\\
\medskip
$\displaystyle
\vcenter{\hbox{\includegraphics[scale=0.75]{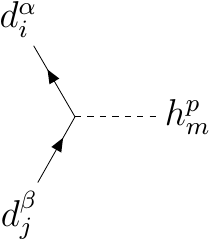}}}
\quad
\frac{1}{\sqrt{2}} \left( \left(\mathcal{W}_d\right)^m_{ij} P_L -
 \left(\mathcal{W}_d^*\right)^m_{ji} P_R \right) \delta_{\alpha\beta}
$
\caption{Feynman rules for the interactions between down-type quarks and
scalars (top) and pseudoscalars (bottom).  The indices $i$ and $j$ label
flavour, $\alpha$ and $\beta$ are color indices, and $m$ indicates the
scalar mass eigenstate.}
\label{fig:FeynmanRules}
\end{figure}

\begin{figure}
\centering
\includegraphics{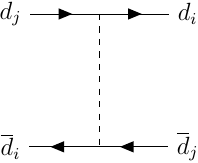} \qquad
\includegraphics{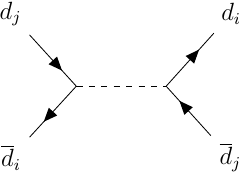}
\caption{Tree-level contributions to FCNC observables from
scalar and pseudoscalar exchange.}
\label{fig:DeltamKScalar}
\end{figure}

\paragraph{Scalar contributions}
From Eqs.~\eqref{eq:H0Decomposition}, \eqref{eq:LQuarkHiggs},
\eqref{eq:diagconvmat}, and \eqref{eq:ScalarMixing},
we obtain the interaction Lagrangian for quarks, scalars, and pseudoscalars
\cite{Escudero:2005hk}
\begin{align}
\mathcal{L} &=
-\!\sum_{l=1,3,5} \frac{1}{\sqrt{2}} \left[
 \left(S_\sigma^T\right)_{lm} h^s_m +
 i \left(S_\varphi^T\right)_{lm} h^p_m
\right] \,
\overline{d_L} \left(V_L^d\, Y^d_l\, V^{d\dagger}_R\right) d_R
\nonumber\\
{}&\hphantom{={}}
-\!\sum_{k=2,4,6} \frac{1}{\sqrt{2}} \left[
 -\left(S_\sigma^T\right)_{km} h^s_m +
 i \left(S_\varphi^T\right)_{km} h^p_m
\right] \,
\overline{u_L} \left(V_L^u\, Y^u_k\, V^{u\dagger}_R\right) u_R
+ \text{h.c.}
\nonumber\\
{}&=
-\frac{1}{\sqrt{2}} \left[
h_m^s \, \overline{d_i} \left(\mathcal{V}_d\right)^m_{ij} P_L \, d_j +
h_m^s \, \overline{d_i} \left(\mathcal{V}_d^*\right)^m_{ji} P_R \, d_j \right.
\nonumber\\
{}&\hphantom{=-\frac{1}{\sqrt{2}} }
\left.{}+
h_m^p \, \overline{d_i} \, i \left(\mathcal{W}_d\right)^m_{ij} P_L \, d_j -
h_m^p \, \overline{d_i} \, i \left(\mathcal{W}_d^*\right)^m_{ji} P_R \,d_j+
\dots \right] ,
\label{eq:ScalaLagMassE}
\end{align}
where $i,j$ denote quark mass eigenstates (flavours) and
$P_{L,R}$ are the chirality projectors.
The dots represent analogous terms for the up-type quarks.
This Lagrangian yields the Feynman rules for down-type quark
interactions in \Figref{fig:FeynmanRules}.
The couplings of scalars and pseudoscalars to down-type quarks are%
\footnote{$\left(\mathcal{V}_d\right)^m$ etc.\ are understood to be
matrices in flavour space if the lower indices are omitted.}
\begin{equation}
	\left(\mathcal{V}_d\right)^m_{ij} =
	\sum_{l=1,3,5}
	\left( V_R^d \, Y^{d\dagger}_l \, V^{d\dagger}_L \right)_{ij}
	\left(S_\sigma\right)_{ml}
\quad,\quad
	\left(\mathcal{W}_d\right)^m_{ij} =
	-\sum_{l=1,3,5}
	\left( V_R^d \, Y^{d\dagger}_l \, V^{d\dagger}_L \right)_{ij}
	\left(S_\varphi\right)_{ml} .
 \label{eq:scalarcoupsME}
\end{equation}
The couplings to up-type quarks are analogous, except that the doublet
indices are summed over $k=2,4,6$ and that there is an extra minus sign
in $\mathcal{V}_u$.
The matrices that diagonalise the quark mass matrices do not diagonalise
the corresponding Yukawa couplings in general.  Hence, both scalar and
pseudoscalar interactions with quarks are expected to violate flavour,
leading to tree-level FCNC and CP violation by scalar and pseudoscalar
exchange, as shown in \Figref{fig:DeltamKScalar}.

\paragraph{Effective Hamiltonian}
We first calculate the amplitude of a $\Delta F = 2$ process ($F=S,B$)
mediated by $Z'$, scalar, and pseudoscalar exchange in the UV-complete
theory (Figs.~\ref{fig:DeltamKVector} and \ref{fig:DeltamKScalar}),
and afterwards match onto the effective Hamiltonian
\begin{equation} \label{eq:effHam1}
	H^{ij}_{\Delta F=2} =
	\big[H^{ij}_{\Delta F=2}\big]_\text{SM} +
	\sum_{r=1}^5 C^{ij}_r(\mu) \, O^{ij}_r +
	\sum_{r=1}^3 \tilde C^{ij}_r(\mu) \, \tilde O^{ij}_r \,,
\end{equation}
where the first term on the right-hand side denotes the contribution
from SM particles and where no sum is implied over the flavour indices.
The Hamiltonian contains the $\Delta F = 2$ operators \cite{Gabbiani:1996hi}
\begin{align} \label{eq:SUSYBasis}
O_1^{ij} &= \overline{d_i^\alpha} \gamma_\mu P_L d_j^\alpha \;
\overline{d_i^\beta} \gamma^\mu P_L
d_j^\beta \,,
\nonumber\\
O_2^{ij} &= \overline{d_i^\alpha} P_L d_j^\alpha \; \overline{d_i^\beta} P_L d_j^\beta
\,,
\nonumber\\
O_3^{ij} &= \overline{d_i^\alpha} P_L d_j^\beta \; \overline{d_i^\beta} P_L d_j^\alpha
\,,
\\
O_4^{ij} &=\overline{d_i^\alpha} P_L d_j^\alpha \; \overline{d_i^\beta} P_R d_j^\beta
\,,
\nonumber\\
O_5^{ij} &= \overline{d_i^\alpha} P_L d_j^\beta \; \overline{d_i^\beta} P_R d_j^\alpha
\,,
\nonumber
\end{align}
where $\alpha$ and $\beta$ are color indices.
The operators $\tilde O_r^{ij}$ are obtained from the corresponding
$O_r^{ij}$ by exchanging $P_L$ and $P_R$.

The calculation of the Feynman diagrams in \Figref{fig:DeltamKVector} gives the $Z'$ contribution to the Wilson coefficients
\begin{align}
    C^{ij}_1(\muEW)&=\frac{g_F^2}{2m_{Z'}^2} (g_L^d)_{ij}^2 \,,
\nonumber\\
    \tilde{C}^{ij}_1(\muEW)&=\frac{g_F^2}{2m_{Z'}^2} (g_R^d)_{ij}^2 \,,
\\
    C^{ij}_5(\muEW)&=-\frac{2 g_F^2}{m_{Z'}^2} (g_L^d)_{ij}(g_R^d)_{ij} \,.
\nonumber
\end{align}
From the diagrams in \Figref{fig:DeltamKScalar}, we obtain
\begin{align}
	C_2^{ij}(\muEW) &= -\frac{1}{4} \left[
	\sum_{m=1}^9 \left(\frac{1}{m^s_m}\right)^2
	\left(\left(\mathcal{V}_d\right)^m_{ij}\right)^2 -
	\sum_{m=3}^9 \left(\frac{1}{m^p_m}\right)^2
	\left(\left(\mathcal{W}_d\right)^m_{ij}\right)^2 \right] ,
\nonumber\\
	\tilde C_2^{ij}(\muEW) &= -\frac{1}{4} \left[
	\sum_{m=1}^9 \left(\frac{1}{m^s_m}\right)^2
	\left(\left(\mathcal{V}_d^*\right)^m_{ji}\right)^2 -
	\sum_{m=3}^9 \left(\frac{1}{m^p_m}\right)^2
	\left(\left(\mathcal{W}_d^*\right)^m_{ji}\right)^2 \right] ,
\\
	C_4^{ij}(\muEW) &= -\frac{1}{2} \left[
	\sum_{m=1}^9 \left(\frac{1}{m^s_m}\right)^2
	\left(\mathcal{V}_d\right)^m_{ij} \left(\mathcal{V}_d^*\right)^m_{ji} +
	\sum_{m=3}^9 \left(\frac{1}{m^p_m}\right)^2
	\left(\mathcal{W}_d\right)^m_{ij} \left(\mathcal{W}_d^*\right)^m_{ji}
	\right] .
\nonumber
\end{align}
The Wilson coefficients $C_3^{ij}$ and $\tilde C_3^{ij}$ are not
generated and thus vanish at the matching scale, which we take to be the
EW scale $\muEW=160\GeV$.
This implies that we neglect the running of the Wilson coefficients between
the different masses of the heavy scalars and pseudoscalars.
While desirable, a precise calculation would have to
compute the running with several intermediate mass scales, making it a
complex task beyond the scope of this work.

We plug the results for the Wilson coefficients into the master formula
for $\Delta F = 2$ transitions valid below the EW scale
\cite{Aebischer:2020dsw},
\begin{equation} \label{eq:Master}
	\big[M_{12}^{ij}\big]_\text{BSM} =
	\frac{1}{2} \, (\Delta M_{ij})_\text{exp} \left[
	\sum_{r=1}^5 P_r^{ij}(\muEW) \, C_r^{ij}(\muEW) +
	\sum_{r=1}^3 P_r^{ij}(\muEW) \, \tilde C_r^{ij}(\muEW) \right] ,
\end{equation}
where $(\Delta M_{ij})_\text{exp}$ are the experimentally measured
values of the neutral meson mass differences and
\begin{equation} \label{eq:DefPij}
	P_r^{ij}(\muEW) =
	\frac{\braket{M^0 | O_r^{ij} | \overline{M^0}}\!(\muEW)}{M_{M^0} \, (\Delta M_{ij})_\text{exp}}
\end{equation}
are the hadronic matrix elements of the $\Delta F=2$ operators
RG-evolved to the EW scale, up
to a normalization factor. Here $M^0$ denotes the relevant
neutral meson.  The numerical values of the $P$ coefficients
are given in \Tabref{tab:Pij}.
\begin{table}
\centering
\renewcommand{\arraystretch}{1.3}
\begin{tabular}{|cc|cccccc|c|}
\hline
System & $ij$ & $P_1^{ij}$ & $P_2^{ij}$ & $P_3^{ij}$ & $P_4^{ij}$ & $P_5^{ij}$ & Units & $(\Delta M_{ij})_\text{exp}$ \\
\hline
$K^0$-$\overline{K^0}$     & $ds$ &  0.102 & $-4.32$ & 1.09 & 14.14 & 4.28 & $10^{13} \GeV^2$ & $3.484 \times 10^{-15} \GeV$ \\
$B^0$-$\overline{B^0}$     & $db$ &  2.67  & $-4.99$ & 1.12 & 12.75 & 5.15 & $10^{11} \GeV^2$ & $3.334 \times 10^{-13} \GeV$ \\
$B^0_s$-$\overline{B^0_s}$ & $sb$ &  1.15  & $-2.24$ & 0.51 &  5.22 & 2.10 & $10^{10} \GeV^2$ & $1.1683 \times 10^{-11} \GeV$ \\
\hline
\end{tabular}
\caption{Numerical values needed for the master formula \eqref{eq:Master} \cite{Aebischer:2020dsw,ParticleDataGroup:2018ovx}.  The $P_r^{ij}$ are given at the scale $\muEW$.
}
\label{tab:Pij}
\end{table}
Note that since QCD conserves P and CP and since we only take into
account the QCD contribution to the RG evolution, the matrix elements
and thus the $P$ coefficients are the same for $O_r$ and $\tilde O_r$.

The quantity $\big[M_{12}^{ij}\big]_\text{BSM}$ contributes to the mass
matrix element
\begin{equation} \label{eq:def-M12}
	M_{12}^{ij} =
	\big[M_{12}^{ij}\big]_\text{SM} + \big[M_{12}^{ij}\big]_\text{BSM} =
	\frac{\braket{M^0 | H^{ij}_{\Delta F=2} | \overline{M^0}}}{2 M_{M^0}}
	\,,
\end{equation}
which determines the observables.  We neglect the contributions of operators of dimension
$8$ or higher, which results in good accuracy for $\epsilon_K$ and
$\Delta m_{B_q}$.

For the neutral kaon system\footnote{The expression for $\epsilon_K$ follows the PDG convention and is only valid in a phase convention where $\phi_2=0$,
corresponding to a real $V_{ud}V_{us}^*$, and in the approximation that
also $\phi_0=0$.  The phase of $\epsilon$, $\arg(\epsilon) \approx
\arctan(-2\Delta m/\Delta\Gamma)$, is determined by non-perturbative QCD
dynamics and is experimentally determined to be about $\pi/4$.} \cite{Aebischer:2020dsw,Workman:2022ynf},
\begin{align}
	\epsilon_K &\simeq
	\frac{e^{i\pi/4}}{\sqrt{2}} \,\frac{\im M_{12}^{ds}}{\Delta M_{ds}} \,.
\label{eq:EpsilonK}
\end{align}

For neutral $B$ mesons \cite{Aebischer:2020dsw},
\begin{equation}
\Delta m_{B_q} = 2 \left| M_{12}^{qb} \right| \,.
\end{equation}

In \Tabref{tbl:FExpObs} we present the values for the flavour
observables that we obtain for the benchmark point presented in
Sec.~\ref{sec:scalarpotential}.
To compare BSM and SM contributions to the $B$ meson mass differences,
we show the quantity $2 \, \big| \big[M_{12}^{qb}\big]_\text{BSM} \big|$.  
Although the total mass difference depends on
$\big| M_{12}^{qb} \big| \neq
 \big| \big[M_{12}^{qb}\big]_\text{SM} \big| +
 \big| \big[M_{12}^{qb}\big]_\text{BSM} \big|$,
we see that the BSM contribution is so small as to be safely within the
limits.
In contrast, the CP violation parameter $\epsilon_K$, for which we show
${|\epsilon_K|}_\text{BSM} =
 \frac{1}{\sqrt{2}} \, \big| \im \big[M_{12}^{ds}\big]_\text{BSM} \big| /
 (\Delta M_{ds})_\text{exp}$
as a rough estimate, receives a significant BSM contribution that
is in tension with the constraints at the $(2 \dots 3)\,\sigma$ level.
The sign of the BSM contribution to $\epsilon_K$ is
well-determined.  At the benchmark point, it is negative.  Changing the model parameters can alter the sign and the size of the contribution.
Interestingly, this contribution is dominated by the operator $O_5$
arising from $Z'$ exchange, despite the very large mass of this boson, about $2.2 \times 10^8\GeV$.
As precise computations of $\Delta m_K$ are difficult due to
unknown long-distance contributions and non-negligible contributions
from dimension-8 operators \cite{Donoghue:1992dd,Aebischer:2020dsw}, we
do not consider this observable but anticipate a significant BSM
contribution to it as well.
This discussion demonstrates that the large masses of the new
particles do not prevent experimental tests of the model.
This is similar to supersymmetric models, where the kaon sector is able to yield constraints even for a large squark mass scale of order $10^4\GeV$ (see, e.g., \cite{Kersten:2012ed}).

\begin{table}
\centering
\begin{tabular}{|c|cl|c|}
\hline
Observable & SM / experimental values & & Model contribution\\
\hline
$\big[\Delta m_{B_d}\big]_\text{SM}$ &
\small{$(0.543\pm 0.029)\ \rm{ps}^{-1}$} & \cite{Lenz:2019lvd} &
\\
& $=(3.57 \pm 0.19)\times 10^{-13}$ GeV & & 
$2 \, \big| \big[M_{12}^{db}\big]_\text{BSM} \big| =$
\\
$(\Delta M_{db})_\text{exp}$ & \small{$(50.65\pm 0.19)\times 10^{10}\ \hbar s^{-1}$} & \cite{Workman:2022ynf} &
\small{$4\times 10^{-16}$ GeV}
\\
& \small{$=(3.33\pm 0.013)\times 10^{-13}$ GeV} & &
\\
\hline
\small{$\big[\Delta m_{B_s}\big]_\text{SM}$} & \small{$(18.77\pm 0.86)\ \rm{ps}^{-1}$} & 
\cite{Lenz:2019lvd} & \\
 & $=(1.235 \pm 0.057)\times 10^{-11}$ GeV & &
$2 \, \big| \big[M_{12}^{sb}\big]_\text{BSM} \big| =$
\\
$(\Delta M_{sb})_\text{exp}$ & \small{$(17.765\pm 0.006)\times 10^{12}\ \hbar s^{-1}$} & \cite{Workman:2022ynf}
& \small{$7\times 10^{-16}$ GeV}
\\
& \small{$=(1.169\pm 0.0004)\times 10^{-11}$ GeV} & &
\\
\hline
${|\epsilon_K|}_\text{SM}$ & 
\small{$\left( 2.170 \pm 0.065_\text{pert.} \right.$} & &
\\
& \small{$\left.{} \pm 0.076_\text{nonpert.} \pm 0.153_\text{param.} \right) \times 10^{-3}$} &
\cite{Brod:2022har} &
${|\epsilon_K|}_\text{BSM} =$
\\
& \small{$=(2.170\pm 0.1828)\times 10^{-3}$} & &
\small{$4 \times 10^{-4}$}
\\
\small{${|\epsilon_K|}_\text{exp}$} & \small{$(2.228 \pm 0.0011) \times 10^{-3}$} & \cite{Brod:2022har} & 
\\
\hline
\end{tabular}
\caption{SM, experimental and model values of flavour observables.}
\label{tbl:FExpObs}
\end{table}

\section{Electroweak Precision Observables \label{sec:T}}

The EW precision tests are a useful way to constrain new physics parameters when the new particles couple to the SM $Z$ and/or $W^\pm$ bosons.
Since in our model the new physics scale is much higher than the EW
scale, the three parameters, $S$, $T$, and $U$~\cite{Peskin:1991sw}
can encapsulate the oblique corrections
at the one-loop level.
The best fit values are~\cite{Workman:2022ynf}
\begin{eqnarray}
S &=& -0.02\pm 0.10 \,, \nonumber\\
T &=& 0.03 \pm 0.12 \,,\nonumber\\
U &=& 0.02 \pm 0.11 \,.
\label{eq:STU_fit}
\end{eqnarray}

Since the Higgs doublets are charged under both the SM gauge group and $U(1)_F$, the $Z$ and $Z'$ gauge bosons mix with each other. Writing the mass parameters as 
\begin{equation}
\mathcal{L} \supset \frac{1}{2} M_Z^2 Z_\mu Z^\mu + \frac{1}{2} M_{Z'}^2 Z'_\mu {Z'}^\mu + \delta M^2 Z_\mu {Z'}^\mu
\,,
\end{equation}
we obtain
\begin{eqnarray}
M_Z^2 &=& \frac{g^2}{4 c_W^2} \sum_{i=1}^6 v_i^2 \,,
\nonumber\\
\delta M^2 &=& -\frac{g g_F}{c_W} \sum_{i=1}^6  Q_{X_i} v_i^2 \,,
\nonumber\\
M_{Z'}^2 &=& g_F^2 \sum_{i=1}^9 Q_{X_i}^2 v_i^2 \,,
\end{eqnarray}
where $Q_{X_i}=(-1)^{i-1} c_{H_i}$ ($i=1,\dots, 6$) and
$Q_{X_{i+6}}=c_{\phi_i}$ ($i=1,2,3$). 
Then the physical masses and the mixing angle are obtained as
\begin{eqnarray}
M_{Z_{1(2)}}^2 &=& \frac{1}{2} \left(M_Z^2 + M_{Z'}^2 \mp \sqrt{(M_Z^2 - M_{Z'}^2)^2+4 (\delta M^2)^2} \right),
\nonumber\\
\tan2 \xi &=& \frac{2 \delta M^2}{M_{Z'}^2-M_Z^2} \,.
\end{eqnarray}
At tree level the $Z-Z'$ mixing induces the $T$ parameter
\begin{equation} \label{eq:Ttree}
\alpha T = \xi^2 \left( \frac{M_{Z_2}^2}{M_{Z_1}^2} -1 \right) ,
\end{equation}
where we have used $\xi \ll 1$.
For the parameter space region we consider, $\alpha M_{Z'}^2 \gg M_Z^2$,
and hence the tree-level $T$ parameter
is negligible.

Instead, the most stringent constraint comes from the loop-induced $T$ parameter~\cite{Grimus:2008nb}.
The multi-Higgs contribution to the $T$ parameter can be found in~\cite{Grimus:2007if,Grimus:2008nb}.
From \cite{Grimus:2007if} we read off
\begin{align}
T &= \frac{1}{16\pi s_W^2 m_W^2}\Bigg\{
\sum_{m=2}^6 \sum_{n=1}^9 \bigg[\sum_{a=1}^6 (S_+)_{ma} (S_\sigma)_{na}\bigg]^2 \, F\big((m^+_m)^2, (m^s_n)^2\big)
\nonumber\\
&+
\sum_{m=2}^6 \sum_{n=3}^9 \bigg[\sum_{a=1}^6 (S_+)_{ma} (S_\varphi)_{na}\bigg]^2 \, F\big((m^+_m)^2, (m^p_n)^2\big)
\nonumber\\
&-
2 \sum_{m=1}^9 \sum_{n=3}^9 \bigg[\sum_{a=1}^6 (S_\sigma)_{ma} (S_\varphi)_{na}\bigg]^2 \, F\big((m^s_m)^2, (m^p_n)^2\big)
\nonumber\\
&+
3 \sum_{m=1}^9 \bigg[\sum_{a=1}^6 (S_\sigma)_{ma} (S_\varphi)_{1a}\bigg]^2
\Big[ F\big(m_Z^2,(m^s_m)^2\big)-F\big(m_W^2,(m^s_m)^2\big) \Big]
\nonumber\\
&-
3 \, \Big[ F(m_Z^2,m_h^2)-F(m_W^2,m_h^2) \Big]
\Bigg\} \,,
\end{align}
where $m^+_m$, $m^s_n$, and $m^p_n$ are
charged, scalar, and pseudoscalar masses, respectively. 
We identify the SM-like scalar $h$ with $h_1^s$, so
$m_h \equiv m_1^s =125\GeV$ for viable parameter space points.
Note that the would-be Nambu-Goldstone bosons $h_{1,2}^p$ (eaten by the
$Z$ and the $Z'$, respectively) and $h_1^+$ (eaten by the $W^+$)
do not contribute, as we can see from the absence of terms including
$m^p_{1,2}$ or $m^+_1$.
The mixing matrices $S_\sigma$, $S_\varphi$, and $S_+$ of the scalars
are defined in eqs.~\eqref{eq:ScalarMixing}.
The loop function is given by
\begin{equation}
F(x,y)=\frac{x+y}{2}-\frac{x y}{x-y}\ln\frac{x}{y}
\end{equation}
and $F(x,x)=0$.

Since the scalar contribution to the $S$ parameter is suppressed
compared to the $T$ parameter~\cite{Grimus:2008nb} we do not consider
$S$ in this paper.
The exotic fermions do not couple to $W^\pm$ and $Z$ and thus do not
contribute to the oblique parameters.

The recent $W$ mass measurement by the CDF collaboration,
$M_W = (80\,433.5 \pm 9.4) \MeV$~\cite{CDF:2022hxs},
shows a $7\sigma$ discrepancy with the SM prediction
$M_W^\text{SM} = (80\,357 \pm 6) \MeV$.
The new physics contribution to $M_W$ is related to the oblique parameters as~\cite{Maksymyk:1993zm}
\begin{equation}
\Delta M_W = -\frac{\alpha M_W^{\rm SM}}{4(c_W^2-s_W^2)}\left(S -2 c_W^2 T -\frac{c_W^2-s_W^2}{2s_W^2}U\right).
\end{equation}
With a sizable value for the $T$ parameter only, we have
\begin{equation}
\Delta M_W \approx 450\, T \MeV \,,
\end{equation}
showing that a positive value $T\approx 0.17 \pm 0.02$ can explain the
$M_W$ anomaly.  We can see that a large enhancement is required compared
with the central value of $T$ in eq.~\eqref{eq:STU_fit}.
For the benchmark point of our model the $T$ parameter is negative and
consistent with eq.~\eqref{eq:STU_fit} at the $2\,\sigma$ level.  Thus,
the CDF measurement cannot be accommodated.
However, this conclusion changes if we accept a smaller mass of order
$1\GeV$ for the lightest pseudoscalar.  In this case, $T$ can be raised
towards the required value by a suitable choice of parameters.

\section{Conclusions \label{sec:conclusions}}
The discovery of the $125\GeV$ Higgs boson
marked the last milestone of the SM as we
know it. But it has left open many questions that seem unfeasible to
answer with the LHC experiments. Among them is the question of how it is possible
that the masses of the SM fermions are so hierarchical if they
couple to the same Higgs field. This question has been tackled in some
ways, among others the Froggatt-Nielsen mechanism
\cite{Froggatt:1978nt}, that allows to fix the hierarchy of Yukawa
couplings through the vacuum expectation values of scalars,
\emph{flavons}. They acquire their vevs at a very high energy, making it
impossible to shed light on them directly. Lowering the scale at which a
flavon can obtain a vev is very challenging due to the
severe constraints from flavour-changing neutral currents (FCNC), but
could offer the possibility of associated signals that could be detected
at low energies. In this work, we have shown that adding more Higgs
fields to the SM can explain the hierarchy of
the fermion masses that we observe. The kind of models that we propose follow the same guiding
principle as the SM, that is, renormalizable tree-level couplings to SM fermions. The
hierarchy of fermion masses is fixed not by powers of the flavon vev as in
Froggatt-Nielsen models, but by allowing fermions to
couple with only a given Higgs field. This is achieved by introducing a
family gauge group $U(1)_F$ and fixing the coupling of fermions and Higgs
fields with suitable charge assignments, respecting the condition of
anomaly cancellation.  This a kind of minimal condition.

The hierarchy of fermion masses and mixings is hence determined by the hierarchy
of Higgs vevs as we assume $\mathcal{O}(1)$ Yukawa couplings. The hierarchy of Higgs vevs is in turn determined from
observational constraints, including the condition to give a spectrum
of scalars consistent with observations, that is, a light scalar of
around $125\GeV$ and heavy scalars with masses beyond the current
bounds. It is well known that tree-level couplings to different Higgs
fields can induce large FCNC and CP violation.
To study this in detail, we have constructed the effective $\Delta F=2$
Lagrangian with vector contributions mediated by the extra
$U(1)_F$ gauge boson and scalar contributions mediated by the
extra Higgs fields. The effective $\Delta F=2$ Lagrangian arising from
scalar exchange has been studied previously in the context of
supersymmetric models \cite{Escudero:2005hk}, but not in the context of
non-supersymmetric models. Besides, the way we tackle how Yukawa
couplings are generated is different from previously considered mechanisms. We have given general expressions for the Wilson coefficients that can be used for further model building with any number of Higgs doublets and scalars.

One salient feature of our model, and in general of the kind of models
we propose, is the appearance of a light pseudoscalar that constrains
some couplings of the theory but could offer an interesting way to
discern it from other theories for the generation of fermion masses and
mixings, which often do not have any physical consequence at low energy. 
 While not constrained by current bounds, this pseudoscalar could be detectable in non-standard Higgs decays or $\Upsilon$ decays in the future.
The mass of this particle is correlated with the electroweak precision
observable $T$ and thus with the $W$ mass.  The best-fit value of $T$
and in particular the larger positive value required to explain the CDF
anomaly appear to favor a smaller pseudoscalar mass.

The model introduced in \cref{sec:specificexample} is not minimal and additional symmetries or different charge
assignments may be able to make it more predictive, but it serves as a
proof of principle that more commonly explored extensions of the SM are
not alone in the quest to explain the masses of the SM particles.

The specific model we have analyzed could lead not only to
FCNC but also to flavour-violating Higgs decays.
The latter phenomena turn out to be strongly suppressed, but 
FCNC and CP violation in the neutral kaon sector in particular appear promising to probe the parameter space, which motivates more precise calculations.
We have found that $\epsilon_K$ receives a significant
BSM contribution, estimated to be in $(2 \dots 3)\,\sigma$ tension with
constraints at the parameter space point we have considered.
This contribution turns out to stem mainly from $Z'$ exchange, despite
the very large mass of this particle of order $10^8\GeV$. 
An important step in further developing these theories is to understand
whether this is a general feature of this class of models and which part
of the parameter space remains viable.
Further investigations are undoubtedly
warranted, especially in view of the fact that Higgs factories will be a
great opportunity to test this kind of theories.
We have briefly mentioned the possibility of getting Gravitational Waves peaking around $0.1$ Hz, that is in the LISA region.

\section*{Acknowledgements}
We would like to thank Per Osland for useful discussions.
S.B.\ is grateful to the Korea Institute for Advanced Study for the
support and warm hospitality shown during his visit. S.B.\ was supported
in part by the National Research Foundation of Korea (NRF) grant funded
by the Korean government (MSIT), Grant No.\ NRF-2018R1A2A3075605 and RS-2023-00270569. 
J.K.\ and L.V.-S.\ acknowledge support from the ``Fundamental Research
Program''  of the Korea Institute for Advanced Study and the warm
hospitality and stimulating environment during the early stages of this work. The
research of L.V.-S.\ was supported by the Basic Science Research Program through the NRF grants NRF-2023R1A2C200536011, NRF-2020R1A6A1A03047877 (CQUeST) and NRF-RS-2023-00273508.
L.V.-S.\ also acknowledges the routes 110A and 110B of the Seoul Transportation Authority for the reliable internet connection and the inspiring atmosphere. 
J.K.\ and P.K.\ were supported by the NRF through Grant No.~NRF-2019R1A2C3005009.

\appendix

\section{Numerical Form of Mass Matrices}
The numerical form of the mass matrices can be worked out from \eq{eq:Mdsq}, using that $H=M^dM^{d\dagger}$ is a Hermitian matrix and the Cholesky decomposition to write it as $H=L^\dagger\, L$, where $L$ is a  upper triangular matrix. Note that the combination $V_\ckm^\dagger \widehat{M^d}$ in \eq{eq:Mdsq} is in fact very close to an upper triangular, hence this form can be obtained by a multiplication of $L$ by a unitary matrix $U$ such that 
\begin{equation}
H=L^\dagger U U^\dagger  L \,.
\end{equation}
Specifically, to obtain the upper triangular matrix of the form \eq{eq:Mdform} we use 
\begin{equation}
U=
\left(
\begin{array}{ccc}
0 & 0 & 1\\ 
0 & 1 & 0\\
1 & 0 & 0
\end{array}
\right) 
\left(
\begin{array}{ccc}
  1   & 0  & 0  \\
  0   & \cos\theta & \sin\theta\\
  0  &  -\sin\theta & \cos\theta
\end{array}
\right)
\left(
\begin{array}{ccc}
  1   &  0 & 0 \\
   0  &  e^{-i\, \varphi_{3,2}}  & 0\\
   0  &  0  & e^{-i\, \varphi_{3,3}}
\end{array}
\right),
\end{equation}
where the first matrix after the equal sign brings the lower triangular
matrix to an upper triangular form, the second matrix sets the element
$M^d_{1,2}$ different from zero and hence achieves the form of \eq{eq:Mdform}. We have some freedom in setting the value of $\theta$, as long as all the coefficients in the matrices $Y^d_{n}$ remain of $\mathcal{O}(1)$. 
This constraint restricts the parameter space for the angle $\theta$.
We checked that all of the observable quantities vary only slightly
under changes of the $\mathcal{O}(1)$ numbers in the matrices $Y^d_n$.
To find the final form of $M^d$ we then have 
\begin{equation}
M^d=L^\dagger U^\dagger \,.
\end{equation}
The numerical values of the CKM matrix, the mass eigenvalues, the
numerical form of $M^d$ and the values of the vevs
are given in \Tabref{tab:numericalMd}. Note that just from the quark sector it is not possible to define uniquely the value of Yukawa couplings  appearing in \eq{eq:Mdform} and 
\eq{eq:formofMd_MA}, since the quantities that are fixed are the combinations $v_nY^f_{jk}$.
\begin{table}
    \centering
    \begin{tabular}{|c|c|}
    \hline
       $V_\ckm$  &  Diag($M^f(M_Z)$)  \\
       \hline
        & {\small{$\left(\begin{array}{ccc}
        m_d & 0 & 0\\
        0 & m_s & 0\\
        0 & 0 &  m_b\end{array}\right)^{\vphantom{2}}$}} \\
        &=\\
    $\begin{array}{c}
    \sin\theta_{12}=0.22500 \pm 0.00067\;,\; \sin\theta_{13}=0.00369 \pm 0.00011
    \\
    \sin\theta_{23}=0.04182^{+0.00085}_{-0.00074}\;,\; \delta=1.144 \pm 0.027
    \end{array}$
        & {\small{$\!\!\!\left(\!\begin{array}{ccc}
        0.0028 & 0 & 0\\
        0 & 0.055 & 0\\
        0 & 0 &  2.85\end{array}\!\right)\!$}}\\
        {\small{
        $\left(\begin{array}{ccc} 
        1 & 0 & 0\\
        0 & c_{23} & s_{23}\\
        0 & -s_{23} & c_{23}
        \end{array} \right)\left(\begin{array}{ccc} 
        c_{13} & 0 & s_{13}e^{-i\, \delta}\\
        0 & 1 & 0 \\
         -s_{13}e^{-i\, \delta} & 0 & c_{13}
        \end{array} \right)\left(\begin{array}{ccc} 
        c_{12}& s_{12} & 0  \\
        -s_{12} & c_{12} &0 \\
        0 & 0 & 1
        \end{array} \right) $}} & {\small{$\left(\!\begin{array}{ccc}
        m_u & 0 & 0\\
        0 & m_c & 0\\
        0 & 0 &  m_t\end{array}\!\right)$}}  \\
        & =\\
         & {\small{$\!\!\left(\!\begin{array}{ccc}
        0.0013 & 0 & 0\\
        0 & 0.63 & 0\\
        0 & 0 &  171.46\end{array}\!\right)\!$}}\\[17pt]
        \hline
        $M^d$ &  $v_f$\\
        \hline
        {\small{$\!\!\left(\!\begin{array}{ccc}
        0 & -0.00238672 -0.0145524\, i & 0.00714217-0.0000167\, i \\
        0  &  0.0561265  -0.0331034\, i & 0.105408 + 0.0291736\, i\\
        0.316899   & 1.87105 & 2.12316
        \end{array}\!\right)\!\!$}} &  
       {\small{$\!\!\!\left(\!\begin{array}{c}
       0.005 \\
       0.05 \\
       1.0
 \end{array}\!\right)^{\vphantom{2}}_{\!d}$}}\;,
       {\small{$\left(\!\begin{array}{c}
       0.005 \\
       0.05 \\
       245.997
 \end{array}\!\right)_{\!u}$}}\!\!
        \\[17pt]
        \hline
    \end{tabular}
    \caption{Numerical values of $V_\ckm$ and $M^d$. All masses and vevs
are given in GeV\@.  The value of $v_6$ has been obtained from $v_6^2=v^2-\sum_{i=1}^5 v_i^2$, $v=246.0\GeV$. \label{tab:numericalMd}}
\end{table}

\section{Benchmark Points \label{app:benchmarkpoints}}
In the following tables,
\Tabref{tab:coeficientsb}--\ref{tab:spectrumBP5}, we specify five different benchmark points that illustrate the possible phenomenology of our class of models.

\begin{table}
\centering
\begin{tabular}{|cc|cc|cc|}
\hline
\multicolumn{6}{|c|}{BP1}\\
\hline
\multicolumn{2}{|c|}{Vevs / couplings} &
\multicolumn{4}{c|}{Quartic couplings} \\
\hline
  $v_7$ & $1.10 \times 10^8 \GeV$ &
  $b_{\phi_1\phi_1\phi_2\phi_3}$ & $-0.400$ & $b_{1221}$ & $4.67$ \\
  $v_8$ & $200 \GeV$ &
  $b_{\phi_1 55}$ & $4.53$ & $b_{1331}$ & $4.51$ \\
  $v_9$ & $1.50 \times 10^3 \GeV$ &
  $b_{\phi_1 66}$ & $-1.10$ & $b_{1441}$ & $3.46$ \\
  $\tilde\mu_{\phi_1 16}$ & $-1.00 \times 10^3 \GeV$ &
  $b_{\phi_1 44}$ & $4.03$ & $b_{1551}$ & $3.12$ \\
  $\tilde\mu_{\phi_2 36}$ & $-10.0 \GeV$ &
  $b_{\phi_1 11}$ & $4.42$ & $b_{1661}$ & $1.20$ \\
  $\tilde\mu'_{\phi_3 23}$ & $5.00 \times 10^8 \GeV$ &
  $b_{\phi_1 22}$ & $1.07$ & $b_{2332}$ & $4.91$ \\
  $\mu_{\phi_1 64}$ & $0.100 \GeV$ &
  $b_{\phi_1 33}$ & $4.11$ & $b_{2442}$ & $0.440$ \\
  $b_{\phi_1\phi_1}$ & $1.18$ &
  $b_{\phi_2 nn} \: \scriptstyle (n=1,\dots,5)$ & $1.00$ & $b_{2552}$ & $2.28$ \\
  $b_{\phi_1\phi_2}$ & $1.31$ &
  $b_{\phi_2 66}$ & $1.00 \times 10^{-5}$ & $b_{2662}$ & $2.17$ \\
  $b_{\phi_1\phi_3}$ & $1.00$ &
  $b_{\phi_3 nn} \: \scriptstyle (n=1,\dots,5)$ & $1.00$ & $b_{3443}$ & $1.51$ \\
  $b_{\phi_2\phi_2}$ & $1.04$ &
  $b_{\phi_3 66}$ & $1.00 \times 10^{-5}$ & $b_{3553}$ & $0.097$ \\
  $b_{\phi_2\phi_3}$ & $1.00$ &
  $b_{1122}$ & $1.43$ & $b_{3663}$ & $0.700$ \\
  $b_{\phi_3\phi_3}$ & $2.41$ &
  $b_{1133}$ & $1.46$ & $b_{4554}$ & $1.74$ \\
  $b_{1111}$ & $3.61$ &
  $b_{1144}$ & $3.47$ & $b_{4664}$ & $0.530$ \\
  $b_{2222}$ & $4.25$ &
  $b_{1155}$ & $1.09$ & $b_{5665}$ & $3.87$ \\
  $b_{3333}$ & $1.70$ &
  $b_{1166}$ & $4.01$ & $\widehat{\tilde{b}}_{3514}$ & $3.64$ \\
  $b_{4444}$ & $3.70$ &
  $b_{2233}$ & $1.23$ & $\widehat{\tilde{b}}_{3154}$ & $0$ \\
  $b_{5555}$ & $1.49$ &
  $b_{2244}$ & $2.71$ & $\widehat{\tilde{b}}_{3415}$ & $0$ \\
  $b_{6666}$ & $0.403$ &
  $b_{2255}$ & $0.550$ & $\widehat{\tilde{b}}_{4616}$ & $2.14$ \\
  & & $b_{2266}$ & $2.67$ & $\tilde{b}'_{\phi_1\phi_1 14}$ & $-1.35$ \\
  & & $b_{3344}$ & $2.27$ & $\tilde{b}'_{\phi_1\phi_2 34}$ & $-7.03$ \\
  & & $b_{3355}$ & $3.27$ & $\tilde{b}_{\phi_1\phi_2 45}$ & $0.100$ \\
  & & $b_{3366}$ & $0.840$ & $b'_{\phi_1\phi_1 53}$ & $-3.00$ \\
  & & $b_{4455}$ & $1.92$ & $\tilde{b}'_{\phi_1\phi_3 25}$ & $4.71$ \\
  & & $b_{4466}$ & $1.02$ & $b'_{\phi_2\phi_3 26}$ & $0.590$ \\
  & & $b_{5566}$ & $0.150$ & $b_{\phi_1 \phi_213}$ & $1.30$ \\
\hline
\end{tabular}
\caption{Numerical values of scalar potential parameters that were used to obtain the spectrum in \Tabref{tab:spectrumA}. The mass parameters $\mu_{\phi_p}^2$ and $\mu_{nn}^2$ are determined by the conditions $\partial V/\partial \sigma_k=0$, where $\sigma_k$ are the real components of the neutral scalar fields, see Eqs.~\eqref{eq:H0Decomposition} and \eqref{eq:PhiDecomposition}. The vevs $v_i$ for $i=1,\dots,6$ are given in \Tabref{tab:numericalMd}.
\label{tab:coeficientsb}}
\end{table}
\begin{table}
    \centering
    \begin{tabular}{|cc|cc|cc|}
    \hline
    \multicolumn{6}{|c|}{BP1}\\
    \hline
    \multicolumn{2}{|c|}{Scalar masses} &
    \multicolumn{2}{c|}{Pseudoscalar masses} &
    \multicolumn{2}{c|}{Charged scalar masses} \\
    \hline
       $m^s_1$ & $125.2 \GeV$ & & & & \\
       $m^s_2$ & $3.15 \times 10^3 \GeV$ & $m^p_3$ & $14.7 \GeV$ & & \\
       $m^s_3$ & $1.56 \times 10^4 \GeV$ & $m^p_4$ & $1.56 \times 10^4 \GeV$ & $m^+_2$ & $1.56 \times 10^4 \GeV$ \\
       $m^s_4$ & $6.02 \times 10^6 \GeV$ & $m^p_5$ & $6.02 \times 10^6 \GeV$ & $m^+_3$ & $6.02 \times 10^6 \GeV$ \\
       $m^s_5$ & $9.11 \times 10^6 \GeV$ & $m^p_6$ & $9.11 \times 10^6 \GeV$ & $m^+_4$ & $9.11 \times 10^6 \GeV$ \\
       $m^s_6$ & $1.36 \times 10^8 \GeV$ & $m^p_7$ & $1.36 \times 10^8 \GeV$ & $m^+_5$ & $2.94 \times 10^8 \GeV$ \\
       $m^s_7$ & $1.69 \times 10^8 \GeV$ & $m^p_8$ & $2.94 \times 10^8 \GeV$ & $m^+_6$ & $6.03 \times 10^8 \GeV$ \\
       $m^s_8$ & $2.94 \times 10^8 \GeV$ & $m^p_9$ & $6.03 \times 10^8 \GeV$ & & \\
       $m^s_9$ & $6.03 \times 10^8 \GeV$ & & & & \\
    \hline
    \end{tabular}
    \caption{Spectrum for the example in \cref{sec:HiggsPheno} (benchmark point 1) with the values of the coefficients in \Tabref{tab:numericalMd} and \Tabref{tab:coeficientsb}.}
    \label{tab:spectrumA}
\end{table}
%
%
\begin{table}
    \centering
    \begin{tabular}{|cc|cc|cc|}
    \hline
    \multicolumn{6}{|c|}{BP2}\\
\hline
\multicolumn{2}{|c|}{Vev/couplings} &
    \multicolumn{2}{c|}{Scalar masses} &
    \multicolumn{2}{c|}{Pseudoscalar masses} \\
\hline
$\tilde\mu'_{\phi_3 23}$ & $7 \times 10^8 \GeV$ & $m^s_1$ & $125.0 \GeV$ & $m^p_1$ & $17.2 \GeV$ \\
\hline
\end{tabular}
\caption{Benchmark point 2. The parameter $\tilde\mu'_{\phi_3 23}$ controls to a great extent the mass of the lightest pseudoscalar. Increasing its value increases the lightest pseudoscalar mass. All of the other parameters are like in \Tabref{tab:coeficientsb}. The rest of the resulting spectrum is like that of \Tabref{tab:spectrumA} and so we do not repeat it here. In this case the $T$ parameter is $-0.21$, within $2\sigma$ of the experimental value.}
\end{table}
\begin{table}
    \centering
    \begin{tabular}{|cc|cc|}
\hline
\multicolumn{4}{|c|}{BP3}\\
\hline
\multicolumn{2}{|c|}{Vevs / couplings} &
\multicolumn{2}{c|}{Quartic couplings} \\
\hline
$v_7$ & $2.1 \times 10^6 \GeV$ &
     $b_{6666}$ & $0.405$ \\
\hline
\end{tabular}
\caption{Benchmark point 3. The parameter $v_7$ has the important role of reducing drastically the mass of the lightest pseudoscalar and at the same time the value of $|T|$, leading to $T=-0.003$ for this case. All of the other parameters are like in \Tabref{tab:coeficientsb}. The spectrum for this case changes with respect to the previous benchmark points and we present it in \Tabref{tab:spectrumBP3}. \label{tb:BP3pars}}
\end{table}
\begin{table}
    \centering
    \begin{tabular}{|cc|cc|cc|}
    \hline
    \multicolumn{6}{|c|}{BP3}\\
    \hline
    \multicolumn{2}{|c|}{Scalar masses} &
    \multicolumn{2}{c|}{Pseudoscalar masses} &
    \multicolumn{2}{c|}{Charged scalar masses} \\
    \hline
       $m^s_1$ & $125.0 \GeV$ & & & & \\
       $m^s_2$ & $2.7 \times 10^3 \GeV$ & $m^p_3$ & $7 \GeV$ & & \\
       $m^s_3$ & $3.16 \times 10^3 \GeV$ & $m^p_4$ & $2.1 \times 10^3 \GeV$ & $m^+_2$ & $2.10 \times 10^3 \GeV$ \\
       $m^s_4$ & $4.42 \times 10^5 \GeV$ & $m^p_5$ & $4.42 \times 10^5 \GeV$ & $m^+_3$ & $4.42 \times 10^5 \GeV$ \\
       $m^s_5$ & $2.47 \times 10^6 \GeV$ & $m^p_6$ & $2.47 \times 10^6 \GeV$ & $m^+_4$ & $2.59 \times 10^6 \GeV$ \\
       $m^s_6$ & $2.59 \times 10^6 \GeV$ & $m^p_7$ & $2.59 \times 10^6 \GeV$ & $m^+_5$ & $9.83 \times 10^6 \GeV$ \\
       $m^s_7$ & $3.07 \times 10^6 \GeV$ & $m^p_8$ & $9.83 \times 10^6 \GeV$ & $m^+_6$ & $1.10 \times 10^7 \GeV$ \\
       $m^s_8$ & $1.10 \times 10^7 \GeV$ & $m^p_9$ & $1.10 \times 10^7 \GeV$ & & \\
       $m^s_9$ & $1.10 \times 10^7 \GeV$ & & & & \\
    \hline
    \end{tabular}
    \caption{Spectrum for the example in \Tabref{tb:BP3pars}.
    \label{tab:spectrumBP3}}
\end{table}
\begin{table}
    \centering
    \begin{tabular}{|cc|cc|cc|}
\hline
\multicolumn{6}{|c|}{BP4}\\
\hline
\multicolumn{2}{|c|}{Vevs / couplings} &
\multicolumn{4}{c|}{Quartic couplings} \\
\hline
$v_7$ & $2.1 \times 10^6 \GeV$ &
     $b_{6666}$ & $0.405$ &  &  \\
$\tilde\mu'_{\phi_3 23}$  & $2.7 \times 10^3$ GeV &  &  &  &  \\
\hline
\end{tabular}
\caption{Benchmark point 4. Changing simultaneously $v_7$ and $\tilde\mu'_{\phi_3 23}$ can lower even to $1 \GeV$ the mass of the lightest pseudoscalar and change the sign of the $T$ parameter. For this case we have $T=0.028$. 
All of the other parameters are like in \Tabref{tab:coeficientsb}. The spectrum for this case changes with respect to the previous benchmark points and we present it in \Tabref{tab:spectrumBP4}. \label{tb:BP4pars}}
\end{table}
\begin{table}
    \centering
    \begin{tabular}{|cc|cc|cc|}
    \hline
    \multicolumn{6}{|c|}{BP4}\\
    \hline
    \multicolumn{2}{|c|}{Scalar masses} &
    \multicolumn{2}{c|}{Pseudoscalar masses} &
    \multicolumn{2}{c|}{Charged scalar masses} \\
    \hline
       $m^s_1$ & $125.0 \GeV$ & & & & \\
       $m^s_2$ & $1.45 \times 10^3 \GeV$ & $m^p_3$ & $1 \GeV$ & & \\
       $m^s_3$ & $3.15 \times 10^3 \GeV$ & $m^p_4$ & $1.45 \times 10^3 \GeV$ & $m^+_2$ & $1.5 \times 10^3 \GeV$ \\
       $m^s_4$ & $2.40 \times 10^5 \GeV$ & $m^p_5$ & $2.4 \times 10^5 \GeV$ & $m^+_3$ & $2.4 \times 10^5 \GeV$ \\
       $m^s_5$ & $8.55 \times 10^5 \GeV$ & $m^p_6$ & $8.55 \times 10^5 \GeV$ & $m^+_4$ & $8.55 \times 10^5 \GeV$ \\
       $m^s_6$ & $1.24 \times 10^6 \GeV$ & $m^p_7$ & $1.24 \times 10^6 \GeV$ & $m^+_5$ & $5.48 \times 10^6 \GeV$ \\
       $m^s_7$ & $1.54 \times 10^6 \GeV$ & $m^p_8$ & $5.48 \times 10^6 \GeV$ & $m^+_6$ & $6.45 \times 10^6 \GeV$ \\
       $m^s_8$ & $5.48 \times 10^6 \GeV$ & $m^p_9$ & $6.45 \times 10^6 \GeV$ & & \\
       $m^s_9$ & $6.45 \times 10^6 \GeV$ & & & & \\
    \hline
    \end{tabular}
    \caption{Spectrum for the example in \Tabref{tb:BP4pars}.
    \label{tab:spectrumBP4}}
\end{table}
\begin{table}
\centering
\begin{tabular}{|cc|cc|}
\hline
\multicolumn{4}{|c|}{BP5}\\
\hline
\multicolumn{2}{|c|}{Vevs / couplings} &
\multicolumn{2}{c|}{Quartic couplings} \\
\hline
$v_7$ & $1 \times 10^6 \GeV$ & $b_{6666}$ & $0.405$ \\
$\tilde\mu'_{\phi_3 23}$ & $3 \times 10^7 \GeV$ & & \\
\hline
\end{tabular}
\caption{Benchmark point 5. We achieve $m_a=3.1\GeV$ for this case. The corresponding $T$ value is $T=0.028$. 
All of the other parameters are like in \Tabref{tab:coeficientsb}. The spectrum for this case changes with respect to the previous benchmark points and we present it in \Tabref{tab:spectrumBP5}. \label{tb:BP5pars}}
\end{table}
\begin{table}
    \centering
    \begin{tabular}{|cc|cc|cc|}
    \hline
    \multicolumn{6}{|c|}{BP5}\\
    \hline
    \multicolumn{2}{|c|}{Scalar masses} &
    \multicolumn{2}{c|}{Pseudoscalar masses} &
    \multicolumn{2}{c|}{Charged scalar masses} \\
    \hline
       $m^s_1$ & $125.0 \GeV$ & & & & \\
       $m^s_2$ & $1.45 \times 10^3 \GeV$ & $m^p_3$ & $3.1 \GeV$ & & \\
       $m^s_3$ & $2.4 \times 10^5 \GeV$ & $m^p_4$ & $1.45 \times 10^3 \GeV$ & $m^+_2$ & $1.5 \times 10^3 \GeV$ \\
       $m^s_4$ & $1.0 \times 10^6 \GeV$ & $m^p_5$ & $2.4 \times 10^5 \GeV$ & $m^+_3$ & $2.4 \times 10^5 \GeV$ \\
       $m^s_5$ & $1.24 \times 10^6 \GeV$ & $m^p_6$ & $1.01 \times 10^6 \GeV$ & $m^+_4$ & $1.01 \times 10^6  \GeV$ \\
       $m^s_6$ & $1.24 \times 10^6 \GeV$ & $m^p_7$ & $1.24 \times 10^6 \GeV$ & $m^+_5$ & $5.48 \times 10^6 \GeV$ \\
       $m^s_7$ & $1.54 \times 10^6 \GeV$ & $m^p_8$ & $5.48 \times 10^6 \GeV$ & $m^+_6$ & $6.45 \times 10^6 \GeV$ \\
       $m^s_8$ & $5.48 \times 10^6 \GeV$ & $m^p_9$ & $6.45 \times 10^6 \GeV$ & & \\
       $m^s_9$ & $6.45 \times 10^6 \GeV$ & & & & \\
    \hline
    \end{tabular}
    \caption{Spectrum for the example in \Tabref{tb:BP5pars}.
    \label{tab:spectrumBP5}}
\end{table}

\section{Charges for Additional Model Building\label{app:additionalch}}
We found that there are fifteen solutions that satisfy the requirements of \eq{eq:PropMd} and \eq{eq:conditionsh1} for the mass matrix of $M^d$ and the conditions of no simultaneous couplings of the Higgs coupling to the $d$ and $u$ sectors. For each of the models presented in \Tabref{tbl:additional_solutions} and \Tabref{tbl:additional_solutions_cont} the Higgs potential will be different and hence the associated specific features that we detail in \cref{sec:specificexample} for the specific model that we present there. 

\begin{table}
\centering
\begin{tabular}{|c|c|cccccll|}
\hline
&{\small{Generation/Charges}}& $ \Fmc_{Q_{L\, i}}$  &$\Fmc_{u_{R\, i}}$ &$ \Fmc_{d_{R\, i}}$ & $\Fmc_{L_{L\, i}}$ & $\Fmc_{e_{R\, i}}$ & $\Fmc_{H_{n}}$ & $\Fmc_{H_{m}}$ \\
\hline
I & {\tiny{$i=1$}} & 3  & -1 & -2  & -3 & -3 &  1 & 4   \\
& {\tiny{$i=2$}} & 0  & 2 &  2 &  3  & 4 &  -2 &  -2  \\
&{\tiny{$i=3$}} & -4 & -5 & 2 &  3  &  5 &  -6 &  1  \\
\hline
II &{\tiny{$i=1$}} & 0 & 2 & -4  & -2 & 1  &  -4 &  -2  \\
&{\tiny{$i=2$}} & 3 & -3 &  4 & 2 & 2 & -1  &  6  \\ 
&{\tiny{$i=3$}} & -5 & -7 & 4 & 6 &9  & -9  &  2 \\
\hline
III &{\tiny{$i=1$}} & -5 & 2 & -2 & -2 & 0 & -7  & -7  \\
&{\tiny{$i=2$}} & 4  & -2 & 2  & -1   &  0 & 2  & 6  \\
&{\tiny{$i=3$}} & 0 & -4 &  2 &  6 & 6 & -2  &  4  \\
\hline
IV &{\tiny{$i=1$}} &-5 & 6 & -2 & -2 & -8  & -7  & -11 \\
&{\tiny{$i=2$}} & 4& -4 &  2 & -1 &  6 &  2 & 8   \\
&{\tiny{$i=3$}} & 0 & -6 &  2 & 6 & 8 & -2  & 6  \\
\hline
V&{\tiny{$i=1$}} & 5 & 1 & -2  &  -6 & -3  & 3 &4  \\
&{\tiny{$i=2$}} & -1& -1 & 2  & 3 & 3 & -3  & 0   \\
&{\tiny{$i=3$}} & -5& -4 & 2  & 6 & 6  & -7  &  -1  \\
\hline
VI &{\tiny{$i=1$}} & 5 &   5& -2 &  0 & -4  & 3  & 0 \\
&{\tiny{$i=2$}} & -1 & -2 & 2  & 1 & 1 & -3  &  1  \\
&{\tiny{$i=3$}} & -5 & -7 & 2  & 2 & 9 & -7  &  2  \\
\hline
VII &{\tiny{$i=1$}} & 5 &  -3 & -2 & -6 & -5  &  3 & 8\\
&{\tiny{$i=2$}} & -1& 3 & 2  &  3& 5 & -3  &  -4  \\
&{\tiny{$i=3$}} & -5 & -4 & 2  & 6 & 6 &  -7 &  -1   \\
\hline
VIII &{\tiny{$i=1$}} & 5 &  -4 & -2 & -2 &  -5 & 3  & 9 \\
&{\tiny{$i=2$}} & -1& 5 & 2  & 2 & 5 & -3  &  -6  \\
&{\tiny{$i=3$}} & -5 & -5 &  2 & 3  & 6  & -7  &  0   \\
\hline
\end{tabular}
\caption{$U(1)_F$ family symmetry charges that we find that satisfy the structure of \eq{eq:PropMd}. 
 All new fermions and scalar singlets have vanishing hypercharge.}
\label{tbl:additional_solutions}
\end{table}

\begin{table}
\centering
\begin{tabular}{|c|c|cccccll|}
\hline
& {\small{Generation/Charges}}& $ \Fmc_{Q_{L\, i}}$  &$\Fmc_{u_{R\, i}}$ &$ \Fmc_{d_{R\, i}}$ & $\Fmc_{L_{L\, i}}$ & $\Fmc_{e_{R\, i}}$ & $\Fmc_{H_{n}}$ & $\Fmc_{H_{m}}$ \\
\hline
IX & {\tiny{$i=1$}} & 5 & -4 & -2 & -7  &  -4 & 3  & 5 \\
&{\tiny{$i=2$}} & -1 & -2 & 2  & 1 &  0 & -3  & 0   \\
&{\tiny{$i=3$}} & -5 & 2 & 2  & 9 & 10 &  -7 &   -2  \\
\hline
X &{\tiny{$i=1$}} & 7 & 6  & -2 & -4  &  -4 & 5  & 1  \\
&{\tiny{$i=2$}} & -2 & -4  & 2  & 3 & 0 &  -4 & 2   \\
&{\tiny{$i=3$}} & -6 & -6 &  2 & 4 & 10 & -8  &  0  \\
\hline
XI &{\tiny{$i=1$}} &7 & 0  & -2 & -10 &  -6 &  5 &7  \\
&{\tiny{$i=2$}} & -2& -1 & 2  & 4 & 4 &  -4 & -1   \\
&{\tiny{$i=3$}} & -6 & -3 & 2  & 9  & 8 &  -8 &  -3  \\
\hline
XII &{\tiny{$i=1$}} & -8 & 2 & -4 & -6 &  0 & -12  & -10 \\
&{\tiny{$i=2$}} &  7& -2 & 4  & 6 & 3  &  3 & 9   \\
&{\tiny{$i=3$}} & -1 & -8 & 4  &6  & 9 & -5  &  7  \\
\hline
XIII &{\tiny{$i=1$}} & -8 &  1 & -4  & -6 & 3  & -12   & -9 \\
&{\tiny{$i=2$}} & 7 & -2 & 4  & 6 & 3 &  3 & 9   \\
&{\tiny{$i=3$}} & -1 & -7 & 4  & 6 & 6 & -5  &  6  \\
\hline
XIV &{\tiny{$i=1$}} & 8 & 4  & -4 & -5 &  0 & 4  & 4 \\
&{\tiny{$i=2$}} & -1& -3 &  4 &  5& 2 & -5  &  2  \\
&{\tiny{$i=3$}} &-9 & -9 &  4 & 5 & 10 & -13  &  0  \\
\hline
XV &{\tiny{$i=1$}} & 8 & -8 & -8  & -10 & 0  &  4 & 9 \\
&{\tiny{$i=2$}} & -1&  0&  4 & 6 & 5  &  -5  &  2 \\
&{\tiny{$i=3$}} & -9 & 0 &  8 & 10 & 7 &  -13  &  -5   \\
\hline
\end{tabular}
\caption{Continuation from \Tabref{tbl:additional_solutions}.}
\label{tbl:additional_solutions_cont}
\end{table} 

\addcontentsline{toc}{section}{References} \frenchspacing
\bibliographystyle{JHEP}
\bibliography{hierarchical_multi_Higgs}

\providecommand{\href}[2]{#2}\begingroup\raggedright\begin{thebibliography}{10}

\bibitem{Ferretti:2006df}
L.~Ferretti, S.F.~King and A.~Romanino, \emph{{Flavour from accidental
  symmetries}},
  \href{https://doi.org/10.1088/1126-6708/2006/11/078}{\emph{JHEP} {\bfseries
  11} (2006) 078} [\href{https://arxiv.org/abs/hep-ph/0609047}{{\ttfamily
  hep-ph/0609047}}].

\bibitem{Babu:1989fg}
K.S.~Babu and E.~Ma, \emph{{Radiative Mechanisms for Generating Quark and
  Lepton Masses: Some Recent Developments}},
  \href{https://doi.org/10.1142/S0217732389002239}{\emph{Mod. Phys. Lett. A}
  {\bfseries 4} (1989) 1975}.

\bibitem{Carena:2004zn}
M.~Carena, A.~Delgado, E.~Ponton, T.M.P.~Tait and C.E.M.~Wagner, \emph{{Warped
  fermions and precision tests}},
  \href{https://doi.org/10.1103/PhysRevD.71.015010}{\emph{Phys. Rev. D}
  {\bfseries 71} (2005) 015010}
  [\href{https://arxiv.org/abs/hep-ph/0410344}{{\ttfamily hep-ph/0410344}}].

\bibitem{Archer:2012qa}
P.R.~Archer, \emph{{The Fermion Mass Hierarchy in Models with Warped Extra
  Dimensions and a Bulk Higgs}},
  \href{https://doi.org/10.1007/JHEP09(2012)095}{\emph{JHEP} {\bfseries 09}
  (2012) 095} [\href{https://arxiv.org/abs/1204.4730}{{\ttfamily 1204.4730}}].

\bibitem{Goertz:2023nii}
F.~Goertz, A.~Pastor-Guti\'errez and J.M.~Pawlowski, \emph{{Flavor hierarchies
  from emergent fundamental partial compositeness}},
  \href{https://doi.org/10.1103/PhysRevD.108.095019}{\emph{Phys. Rev. D}
  {\bfseries 108} (2023) 095019}
  [\href{https://arxiv.org/abs/2307.11148}{{\ttfamily 2307.11148}}].

\bibitem{Patel:2017pct}
K.M.~Patel, \emph{{Clockwork mechanism for flavor hierarchies}},
  \href{https://doi.org/10.1103/PhysRevD.96.115013}{\emph{Phys. Rev. D}
  {\bfseries 96} (2017) 115013}
  [\href{https://arxiv.org/abs/1711.05393}{{\ttfamily 1711.05393}}].

\bibitem{King:2020qaj}
S.J.D.~King and S.F.~King, \emph{{Fermion mass hierarchies from modular
  symmetry}}, \href{https://doi.org/10.1007/JHEP09(2020)043}{\emph{JHEP}
  {\bfseries 09} (2020) 043}
  [\href{https://arxiv.org/abs/2002.00969}{{\ttfamily 2002.00969}}].

\bibitem{Ding:2022bzs}
G.-J.~Ding, S.F.~King, J.-N.~Lu and B.-Y.~Qu, \emph{{Leptogenesis in SO(10)
  models with A$_{4}$ modular symmetry}},
  \href{https://doi.org/10.1007/JHEP10(2022)071}{\emph{JHEP} {\bfseries 10}
  (2022) 071} [\href{https://arxiv.org/abs/2206.14675}{{\ttfamily
  2206.14675}}].

\bibitem{Ding:2022aoe}
G.-J.~Ding, F.R.~Joaquim and J.-N.~Lu, \emph{{Texture-zero patterns of lepton
  mass matrices from modular symmetry}},
  \href{https://doi.org/10.1007/JHEP03(2023)141}{\emph{JHEP} {\bfseries 03}
  (2023) 141} [\href{https://arxiv.org/abs/2211.08136}{{\ttfamily
  2211.08136}}].

\bibitem{Froggatt:1978nt}
C.D.~Froggatt and H.B.~Nielsen, \emph{{Hierarchy of Quark Masses, Cabibbo
  Angles and CP Violation}},
  \href{https://doi.org/10.1016/0550-3213(79)90316-X}{\emph{Nucl. Phys.}
  {\bfseries B147} (1979) 277}.

\bibitem{Tsumura:2009yf}
K.~Tsumura and L.~Velasco-Sevilla, \emph{{Phenomenology of flavon fields at the
  LHC}}, \href{https://doi.org/10.1103/PhysRevD.81.036012}{\emph{Phys. Rev.}
  {\bfseries D81} (2010) 036012}
  [\href{https://arxiv.org/abs/0911.2149}{{\ttfamily 0911.2149}}].

\bibitem{Berger:2014gga}
E.L.~Berger, S.B.~Giddings, H.~Wang and H.~Zhang, \emph{{Higgs-flavon mixing
  and LHC phenomenology in a simplified model of broken flavor symmetry}},
  \href{https://doi.org/10.1103/PhysRevD.90.076004}{\emph{Phys. Rev. D}
  {\bfseries 90} (2014) 076004}
  [\href{https://arxiv.org/abs/1406.6054}{{\ttfamily 1406.6054}}].

\bibitem{Bauer:2016rxs}
M.~Bauer, T.~Schell and T.~Plehn, \emph{{Hunting the Flavon}},
  \href{https://doi.org/10.1103/PhysRevD.94.056003}{\emph{Phys. Rev.}
  {\bfseries D94} (2016) 056003}
  [\href{https://arxiv.org/abs/1603.06950}{{\ttfamily 1603.06950}}].

\bibitem{Khoze:2017tjt}
V.V.~Khoze and M.~Spannowsky, \emph{{Higgsplosion: Solving the hierarchy
  problem via rapid decays of heavy states into multiple Higgs bosons}},
  \href{https://doi.org/10.1016/j.nuclphysb.2017.11.002}{\emph{Nucl. Phys. B}
  {\bfseries 926} (2018) 95}
  [\href{https://arxiv.org/abs/1704.03447}{{\ttfamily 1704.03447}}].

\bibitem{Li:2019bcr}
T.-Q.~Li and C.-X.~Yue, \emph{{Flavons and LFV decays and productions of
  pseudoscalar mesons}},
  \href{https://doi.org/10.1142/S0217732319502882}{\emph{Mod. Phys. Lett. A}
  {\bfseries 34} (2019) 1950288}.

\bibitem{King:2020mau}
S.J.D.~King, S.F.~King, S.~Moretti and S.J.~Rowley, \emph{{Discovering the
  origin of Yukawa couplings at the LHC with a singlet Higgs and vector-like
  quarks}}, \href{https://doi.org/10.1007/JHEP05(2021)144}{\emph{JHEP}
  {\bfseries 21} (2020) 144}
  [\href{https://arxiv.org/abs/2102.06091}{{\ttfamily 2102.06091}}].

\bibitem{Koivunen:2023led}
N.~Koivunen and M.~Raidal, \emph{{Production and decays of 146 GeV flavons into
  $e\mu$ final state at the LHC}},
  \href{https://arxiv.org/abs/2305.00014}{{\ttfamily 2305.00014}}.

\bibitem{Escudero:2005hk}
N.~Escudero, C.~Munoz and A.M.~Teixeira, \emph{{FCNCs in supersymmetric
  multi-Higgs doublet models}},
  \href{https://doi.org/10.1103/PhysRevD.73.055015}{\emph{Phys. Rev.}
  {\bfseries D73} (2006) 055015}
  [\href{https://arxiv.org/abs/hep-ph/0512046}{{\ttfamily hep-ph/0512046}}].

\bibitem{Escudero:2005ku}
N.~Escudero, C.~Munoz and A.M.~Teixeira, \emph{{Phenomenological viability of
  orbifold models with three Higgs families}},
  \href{https://doi.org/10.1088/1126-6708/2006/07/041}{\emph{JHEP} {\bfseries
  07} (2006) 041} [\href{https://arxiv.org/abs/hep-ph/0512301}{{\ttfamily
  hep-ph/0512301}}].

\bibitem{Hill:2019cce}
C.T.~Hill, P.A.~Machado, A.E.~Thomsen and J.~Turner, \emph{{Where are the Next
  Higgs Bosons?}},
  \href{https://doi.org/10.1103/PhysRevD.100.015051}{\emph{Phys. Rev. D}
  {\bfseries 100} (2019) 015051}
  [\href{https://arxiv.org/abs/1904.04257}{{\ttfamily 1904.04257}}].

\bibitem{Hill:2019ldq}
C.T.~Hill, P.A.N.~Machado, A.E.~Thomsen and J.~Turner, \emph{{Scalar
  Democracy}}, \href{https://doi.org/10.1103/PhysRevD.100.015015}{\emph{Phys.
  Rev. D} {\bfseries 100} (2019) 015015}
  [\href{https://arxiv.org/abs/1902.07214}{{\ttfamily 1902.07214}}].

\bibitem{Altmannshofer:2021hfu}
W.~Altmannshofer, S.A.~Gadam, S.~Gori and N.~Hamer, \emph{{Explaining
  $(g-2)_\mu$ with multi-TeV sleptons}},
  \href{https://doi.org/10.1007/JHEP07(2021)118}{\emph{JHEP} {\bfseries 07}
  (2021) 118} [\href{https://arxiv.org/abs/2104.08293}{{\ttfamily
  2104.08293}}].

\bibitem{Porto:2007ed}
R.A.~Porto and A.~Zee, \emph{{The Private Higgs}},
  \href{https://doi.org/10.1016/j.physletb.2008.08.001}{\emph{Phys. Lett. B}
  {\bfseries 666} (2008) 491}
  [\href{https://arxiv.org/abs/0712.0448}{{\ttfamily 0712.0448}}].

\bibitem{Porto:2008hb}
R.A.~Porto and A.~Zee, \emph{{Neutrino Mixing and the Private Higgs}},
  \href{https://doi.org/10.1103/PhysRevD.79.013003}{\emph{Phys. Rev. D}
  {\bfseries 79} (2009) 013003}
  [\href{https://arxiv.org/abs/0807.0612}{{\ttfamily 0807.0612}}].

\bibitem{BenTov:2012xp}
Y.~BenTov and A.~Zee, \emph{{Lepton Private Higgs and the discrete group
  $\Sigma(81)$}},
  \href{https://doi.org/10.1016/j.nuclphysb.2013.03.002}{\emph{Nucl. Phys. B}
  {\bfseries 871} (2013) 452}
  [\href{https://arxiv.org/abs/1202.4234}{{\ttfamily 1202.4234}}].

\bibitem{BenTov:2012cx}
Y.~BenTov and A.~Zee, \emph{{Private Higgs at the LHC}},
  \href{https://doi.org/10.1142/S0217751X13501492}{\emph{Int. J. Mod. Phys. A}
  {\bfseries 28} (2013) 1350149}
  [\href{https://arxiv.org/abs/1207.0467}{{\ttfamily 1207.0467}}].

\bibitem{Fritzsch:1977vd}
H.~Fritzsch, \emph{{Weak-interaction mixing in the six-quark theory}},
  \href{https://doi.org/10.1016/0370-2693(78)90524-5}{\emph{Phys. Lett. B}
  {\bfseries 73} (1978) 317}.

\bibitem{CDF:2022hxs}
{\scshape CDF} collaboration, \emph{{High-precision measurement of the W boson
  mass with the CDF II detector}},
  \href{https://doi.org/10.1126/science.abk1781}{\emph{Science} {\bfseries 376}
  (2022) 170}.

\bibitem{Grimus:2007if}
W.~Grimus, L.~Lavoura, O.M.~Ogreid and P.~Osland, \emph{{A Precision constraint
  on multi-Higgs-doublet models}},
  \href{https://doi.org/10.1088/0954-3899/35/7/075001}{\emph{J. Phys. G}
  {\bfseries 35} (2008) 075001}
  [\href{https://arxiv.org/abs/0711.4022}{{\ttfamily 0711.4022}}].

\bibitem{Ko:2012hd}
P.~Ko, Y.~Omura and C.~Yu, \emph{{A Resolution of the Flavor Problem of Two
  Higgs Doublet Models with an Extra $U(1)_H$ Symmetry for Higgs Flavor}},
  \href{https://doi.org/10.1016/j.physletb.2012.09.019}{\emph{Phys. Lett. B}
  {\bfseries 717} (2012) 202}
  [\href{https://arxiv.org/abs/1204.4588}{{\ttfamily 1204.4588}}].

\bibitem{Allanach:2018vjg}
B.C.~Allanach, J.~Davighi and S.~Melville, \emph{{An Anomaly-free Atlas:
  charting the space of flavour-dependent gauged $U(1)$ extensions of the
  Standard Model}}, \href{https://doi.org/10.1007/JHEP02(2019)082}{\emph{JHEP}
  {\bfseries 02} (2019) 082}
  [\href{https://arxiv.org/abs/1812.04602}{{\ttfamily 1812.04602}}].

\bibitem{Jain:1994hd}
V.~Jain and R.~Shrock, \emph{{Models of fermion mass matrices based on a
  flavor- and generation-dependent U(1) gauge symmetry}},
  \href{https://doi.org/10.1016/0370-2693(95)00472-W}{\emph{Phys. Lett. B}
  {\bfseries 352} (1995) 83}
  [\href{https://arxiv.org/abs/hep-ph/9412367}{{\ttfamily hep-ph/9412367}}].

\bibitem{Ellwanger:2009dp}
U.~Ellwanger, C.~Hugonie and A.M.~Teixeira, \emph{{The Next-to-Minimal
  Supersymmetric Standard Model}},
  \href{https://doi.org/10.1016/j.physrep.2010.07.001}{\emph{Phys. Rept.}
  {\bfseries 496} (2010) 1} [\href{https://arxiv.org/abs/0910.1785}{{\ttfamily
  0910.1785}}].

\bibitem{Cepeda:2021rql}
M.~Cepeda, S.~Gori, V.I.~Martinez~Outschoorn and J.~Shelton, \emph{{Exotic
  Higgs Decays}},
  \href{https://doi.org/10.1146/annurev-nucl-102319-024147}{\emph{Ann. Rev.
  Nucl. Part. Sci.} {\bfseries 72} (2022) 119}
  [\href{https://arxiv.org/abs/2111.12751}{{\ttfamily 2111.12751}}].

\bibitem{Wilczek:1977zn}
F.~Wilczek, \emph{{Decays of Heavy Vector Mesons into Higgs Particles}},
  \href{https://doi.org/10.1103/PhysRevLett.39.1304}{\emph{Phys. Rev. Lett.}
  {\bfseries 39} (1977) 1304}.

\bibitem{Vysotsky:1980cz}
M.I.~Vysotsky, \emph{Strong interaction corrections to semiweak decays:
  {C}alculation of the {$V \to H \gamma$} decay rate to order $\alpha_s$},
  \href{https://doi.org/10.1016/0370-2693(80)90571-7}{\emph{Phys. Lett. B}
  {\bfseries 97} (1980) 159}.

\bibitem{Nason:1986tr}
P.~Nason, \emph{{QCD} radiative corrections to upsilon decay into scalar plus
  gamma and pseudoscalar plus gamma},
  \href{https://doi.org/10.1016/0370-2693(86)90721-5}{\emph{Phys. Lett. B}
  {\bfseries 175} (1986) 223}.

\bibitem{Workman:2022ynf}
{\scshape Particle Data Group} collaboration, \emph{{Review of Particle
  Physics}}, \href{https://doi.org/10.1093/ptep/ptac097}{\emph{PTEP} {\bfseries
  2022} (2022) 083C01}.

\bibitem{Kamenik:2023ytu}
J.F.~Kamenik, A.~Korajac, M.~Szewc, M.~Tammaro and J.~Zupan,
  \emph{{Flavor-violating Higgs and $Z$ boson decays at a future circular
  lepton collider}},
  \href{https://doi.org/10.1103/PhysRevD.109.L011301}{\emph{Phys. Rev. D}
  {\bfseries 109} (2024) L011301}
  [\href{https://arxiv.org/abs/2306.17520}{{\ttfamily 2306.17520}}].

\bibitem{Linde:1990flp}
A.D.~Linde, \emph{Particle physics and inflationary cosmology},
  \href{https://arxiv.org/abs/hep-th/0503203}{{\ttfamily hep-th/0503203}}.

\bibitem{LISACosmologyWorkingGroup:2022jok}
{\scshape LISA Cosmology Working Group} collaboration, \emph{{Cosmology with
  the Laser Interferometer Space Antenna}},
  \href{https://doi.org/10.1007/s41114-023-00045-2}{\emph{Living Rev. Rel.}
  {\bfseries 26} (2023) 5} [\href{https://arxiv.org/abs/2204.05434}{{\ttfamily
  2204.05434}}].

\bibitem{Ko:2011di}
P.~Ko, Y.~Omura and C.~Yu, \emph{{Chiral $U(1)$ flavor models and flavored
  Higgs doublets: the top FB asymmetry and the $Wjj$}},
  \href{https://doi.org/10.1007/JHEP01(2012)147}{\emph{JHEP} {\bfseries 01}
  (2012) 147} [\href{https://arxiv.org/abs/1108.4005}{{\ttfamily 1108.4005}}].

\bibitem{Gabbiani:1996hi}
F.~Gabbiani, E.~Gabrielli, A.~Masiero and L.~Silvestrini, \emph{{A Complete
  analysis of FCNC and CP constraints in general SUSY extensions of the
  standard model}},
  \href{https://doi.org/10.1016/0550-3213(96)00390-2}{\emph{Nucl. Phys. B}
  {\bfseries 477} (1996) 321}
  [\href{https://arxiv.org/abs/hep-ph/9604387}{{\ttfamily hep-ph/9604387}}].

\bibitem{Aebischer:2020dsw}
J.~Aebischer, C.~Bobeth, A.J.~Buras and J.~Kumar, \emph{{SMEFT ATLAS of
  $\Delta$F = 2 transitions}},
  \href{https://doi.org/10.1007/JHEP12(2020)187}{\emph{JHEP} {\bfseries 12}
  (2020) 187} [\href{https://arxiv.org/abs/2009.07276}{{\ttfamily
  2009.07276}}].

\bibitem{ParticleDataGroup:2018ovx}
{\scshape Particle Data Group} collaboration, \emph{{Review of Particle
  Physics}}, \href{https://doi.org/10.1103/PhysRevD.98.030001}{\emph{Phys. Rev.
  D} {\bfseries 98} (2018) 030001}.

\bibitem{Donoghue:1992dd}
J.F.~Donoghue, E.~Golowich and B.R.~Holstein, \emph{{Dynamics of the Standard
  Model}}, Cambridge University Press (2023),
  \href{https://doi.org/10.1017/9781009291033}{10.1017/9781009291033}.

\bibitem{Kersten:2012ed}
J.~Kersten and L.~Velasco-Sevilla, \emph{{Flavour constraints on scenarios with
  two or three heavy squark generations}},
  \href{https://doi.org/10.1140/epjc/s10052-013-2405-y}{\emph{Eur. Phys. J. C}
  {\bfseries 73} (2013) 2405}
  [\href{https://arxiv.org/abs/1207.3016}{{\ttfamily 1207.3016}}].

\bibitem{Lenz:2019lvd}
A.~Lenz and G.~Tetlalmatzi-Xolocotzi, \emph{{Model-independent bounds on new
  physics effects in non-leptonic tree-level decays of B-mesons}},
  \href{https://doi.org/10.1007/JHEP07(2020)177}{\emph{JHEP} {\bfseries 07}
  (2020) 177} [\href{https://arxiv.org/abs/1912.07621}{{\ttfamily
  1912.07621}}].

\bibitem{Brod:2022har}
J.~Brod, S.~Kvedaraite, Z.~Polonsky and A.~Youssef, \emph{{Electroweak
  corrections to the Charm-Top-Quark Contribution to $\epsilon_K$}},
  \href{https://doi.org/10.1007/JHEP12(2022)014}{\emph{JHEP} {\bfseries 12}
  (2022) 014} [\href{https://arxiv.org/abs/2207.07669}{{\ttfamily
  2207.07669}}].

\bibitem{Peskin:1991sw}
M.E.~Peskin and T.~Takeuchi, \emph{{Estimation of oblique electroweak
  corrections}}, \href{https://doi.org/10.1103/PhysRevD.46.381}{\emph{Phys.
  Rev. D} {\bfseries 46} (1992) 381}.

\bibitem{Grimus:2008nb}
W.~Grimus, L.~Lavoura, O.M.~Ogreid and P.~Osland, \emph{{The oblique parameters
  in multi-Higgs-doublet models}},
  \href{https://doi.org/10.1016/j.nuclphysb.2008.04.019}{\emph{Nucl. Phys. B}
  {\bfseries 801} (2008) 81} [\href{https://arxiv.org/abs/0802.4353}{{\ttfamily
  0802.4353}}].

\bibitem{Maksymyk:1993zm}
I.~Maksymyk, C.P.~Burgess and D.~London, \emph{{Beyond S, T and U}},
  \href{https://doi.org/10.1103/PhysRevD.50.529}{\emph{Phys. Rev. D} {\bfseries
  50} (1994) 529} [\href{https://arxiv.org/abs/hep-ph/9306267}{{\ttfamily
  hep-ph/9306267}}].

\end{thebibliography}\endgroup

\end{document}